\documentclass[10pt]{iopart}

\usepackage[english]{babel}
\usepackage{iopams} 
\usepackage{graphicx}
\usepackage{float} 
\usepackage[dvipsnames]{xcolor}
\usepackage[colorlinks,citecolor=blue,linkcolor=black,urlcolor=blue]{hyperref}
\usepackage{iopams}  

\usepackage{subcaption}
\usepackage[font=footnotesize]{caption}
\usepackage{tikz}
\usepackage{pgfplots}
\usepackage{placeins}  
\usepackage{dblfloatfix} 
\usepackage{amsmath_new}
\usepackage[dvipsnames]{xcolor} 
\usepackage{soul}
\usepackage{marginnote} 

\begin{document}

\title[Influence of the density gradient on turbulent heat transport at ion-scales]{Influence of the density gradient on turbulent heat transport at ion-scales: an inter-machine study with the gyrokinetic code \texttt{stella}}

\author{H Thienpondt$^1$, J M Garc\'ia-Rega\~na$^1$, I Calvo$^1$ \\
G Acton$^2$ and M Barnes$^2$}

\address{
	$^1$ Laboratorio Nacional de Fusi\'on, CIEMAT, 28040 Madrid, Spain \\
	$^2$ Rudolf Peierls Centre for Theoretical Physics, University of Oxford, Oxford OX1 3NP, United Kingdom}

\ead{Hanne.Thienpondt@ciemat.es}
\vspace{10pt}

\begin{indented}
\item[]March 2024
\end{indented}

\begin{abstract} 
Efficient control of turbulent heat transport is crucial for magnetic confinement fusion reactors. This work discusses the complex interplay between density gradients and micro-instabilities, shedding light on their impact on turbulent heat transport in different fusion devices. In particular, the influence of density gradients on turbulent heat transport is investigated through an extensive inter-machine study, including various stellarators such as W7-X, LHD, TJ-II and NCSX, along with the Asdex-Upgrade tokamak and the tokamak geometry of the Cyclone Base Case (CBC). Linear and nonlinear simulations are performed employing the $\delta f$-gyrokinetic code \texttt{stella} across a wide range of parameters to explore the effects of density gradients, temperature gradients, and kinetic electrons. A strong reduction in ion heat flux with increasing density gradients is found in NCSX and W7-X due to the stabilization of temperature-gradient-driven modes without significantly destabilizing density-gradient-driven modes. In contrast, the tokamaks exhibit an increase in ion heat flux with density gradients. Notably, the behavior of ion heat fluxes in stellarators does not align with that of linear growth rates. Additionally, this study provides physical insights into the micro-instabilities, emphasizing the dominance of trapped-electron-modes in CBC, AUG, TJ-II, LHD and NCSX, while both the trapped-electron-mode and the passing-particle-driven universal instability contribute significantly in W7-X. 
\end{abstract}  

\vspace{2pc}

\submitto{\NF}


\ioptwocol


\section{Introduction}

To achieve net energy in magnetic confinement fusion reactors, it is important to understand, predict and to some extent control heat transport, and in particular transport caused by turbulence. Experimental observations in both tokamaks \mbox{\cite{wolfe1986effect, kaufmann1988pellet, tubbing1991h}} and stellarators \cite{stroth1998high, komori2006overview, bozhenkov2020high} indicate that peaked density profiles result in a reduction of turbulent heat transport. This reduction is believed to be related to the stabilization of the ion-temperature-gradient (ITG) and electron-temperature-gradient (ETG) driven modes by large density gradients. However, an increasing density gradient eventually destabilizes density-gradient-driven modes such as trapped-electron-modes (TEM). Hence, the question arises whether there exists a specific range of density gradients wherein temperature-gradient-driven modes are stabilized, without significantly destabilizing the density-gradient-driven modes.

To assess the impact of density gradients on turbulent heat transport, simulations are conducted with the $\delta\hspace{-0.1mm}f$-gyrokinetic code \texttt{stella} \cite{barnes2019stella}.  A comprehensive comparative study of different devices is carried out, including different families of stellarators, specifically, the Wendelstein 7-X  (W7-X) stellarator, the Large Helical Device (LHD), the TJ-II stellarator and the National Compact Stellarator Experiment (NCSX), which represent helias, heliotron, heliac and quasi-axisymmetric configurations, respectively. Additionally, the Axially Symmetric Divertor Experiment or ASDEX Upgrade (AUG) tokamak and the tokamak geometry of the Cyclone Base Case (CBC) \cite{dimits2000simulation}, are included.
Previous studies including cross-device comparisons are limited to linear results \cite{rewoldt2005comparison, proll2013collisionless}, or only consider a few points in parameter space for nonlinear simulations \cite{mckinney2019comparison, wang2020global, helander2015advances, sanchez2023instabilities}. In contrast, this work contains both linear and nonlinear results covering a wide range of parameters; the effect of the presence of electrons with and without a temperature gradient, and a study of the velocity dependence of the turbulent part of the distribution function. 

The paper is divided into the following sections. First,  the simulation domain and the coordinates employed by the gyrokinetic code \texttt{stella} are explained in section \ref{sec:stella}. Next, section  \ref{sec:devices} introduces and analyzes the different magnetic configurations, and in \mbox{section \ref{sec:localization}} the localization of the micro-instabilities is inferred from the regions of bad curvature and magnetic wells. 
In \mbox{section \ref{sec:linear}} a comprehensive linear stability analysis of the six considered devices is carried out. The growth rates and frequencies of the micro-instabilities are discussed in section \ref{sec:gammavskxky}, for specific values of the profile gradients and considering a wide range of radial and binormal wavenumbers. Section \ref{sec:linearlineallkys} explores the evolution of the instabilities with increasing density gradients, revealing that the most unstable mode transitions continuously from an ion-temperature-gradient driven mode to a predominantly density-gradient-driven mode in CBC, AUG, TJ-II and LHD. In contrast, NCSX and W7-X display distinct frequency branches, indicating that different micro-instabilities can be distinguished from one another. In section \ref{sec:mostunstable} the most unstable mode for the considered binormal wavenumbers is identified, showing that the linear growth rates increase with the density gradient in CBC, AUG, TJ-II and LHD---and to a lesser extent in NCSX---whereas a modest reduction of the growth rates is found for a small range of density gradients in \mbox{W7-X}. The contributions of the specific plasma gradients are isolated in \mbox{section \ref{sec:linearlineITGTEM}}, which reveals that in NCSX and \mbox{W7-X} there is a range of density gradients for which the ion-temperature-gradient driven modes are stabilized, without driving the density-gradient-driven modes particularly unstable yet. To conclude the linear analysis, in section \ref{sec:gammavsfprimtiprim} the ion temperature and density gradient are scanned across a wide range of parameters. 

Section \ref{sec:nonlinear} is the most important part of this paper, as it investigates the effect of the density gradient on the turbulent heat transport by means of nonlinear gyrokinetic simulations. Section \ref{sec:heat} highlights distinct behaviors of the ion heat flux in different devices. Specifically, in the tokamaks, the ion heat flux increases strongly with the density gradient, whereas in LHD it remains largely unaffected, and in TJ-II a substantial reduction of the ion heat flux occurs for small density gradients. 
In contrast, both NCSX and W7-X exhibit a strong reduction of the ion heat flux across a wide range of density gradients, due to the stabilization of temperature-gradient-driven modes, without significantly destabilizing  density-gradient-driven modes, as demonstrated in \mbox{section \ref{sec:nonlinearITGTEM}}. Additionally, the influence of a finite electron temperature gradient is discussed in \mbox{section \ref{sec:nonlinearWithALTe}}. It is worth noting that in stellarators the ion heat flux shows no correlation with the growth rates obtained from the linear analysis.
Finally, the distribution function is examined in detail in section \ref{sec:distributon} to characterize the micro-instabilities that are driving the turbulence, by means of novel diagnostics developed in \texttt{stella}. Specifically, the velocity dependence of the perturbed distribution function is examined to quantify the importance of the ion and electron dynamics, and to determine whether the turbulence is primarily driven by the trapped or the passing electrons. It is found that at high density gradients, trapped-electron-modes predominately drive the heat transport in CBC, AUG, TJ-II, LHD and NCSX, whereas in W7-X, the passing-particle-driven universal instability contributes significantly to the turbulence.

\begin{figure}[!b]
	\centering 
	\includegraphics[trim={0mm 2mm 0mm 1mm}]{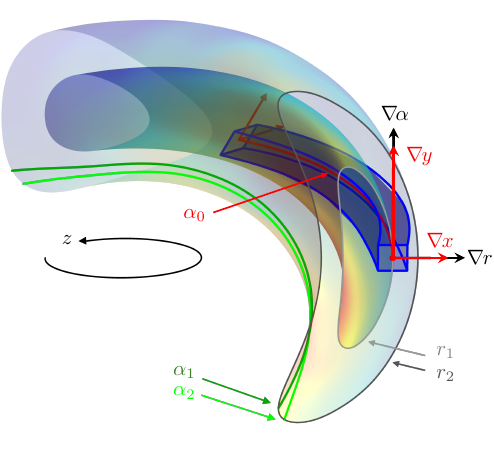}
	\caption{Half a period of the Wendelstein 7-X stellarator depicting the coordinates used in \texttt{stella}. The last closed flux surface and the flux surface located at $r/a = 0.5$ are shown, highlighting two field lines in green. The flux tube surrounding the $\alpha_0=0$ field line (passing through $\theta=0$ and  $\zeta=0$) at $r/a = 0.5$ is shown in blue, extending a quarter of a period.  }
	\label{fig:geometry} 
\end{figure} 

\begin{figure*}[b!]
	\centering
	\includegraphics[trim={0mm 2mm 0mm 1mm}] {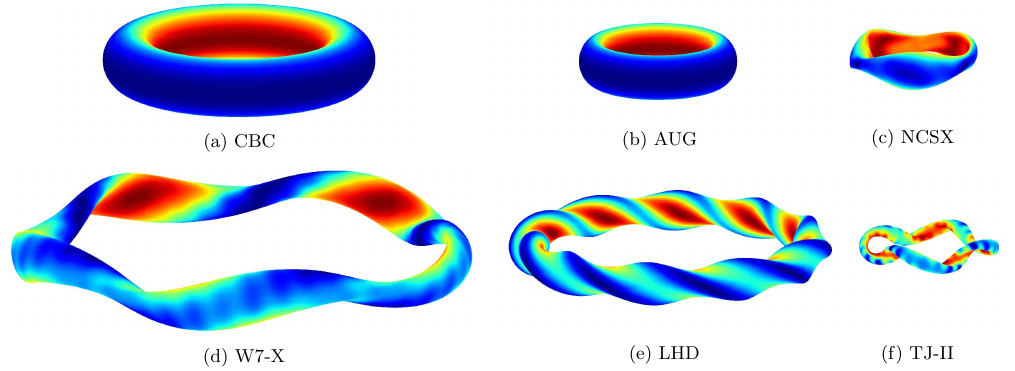}
	
	\caption{Magnetic field strength---minimum is blue, maximum is red---on the flux surface $r/a=0.7$ for the six considered devices. The same scale is used in order to compare the relative size of the reactors.}
	\label{fig:3D}
\end{figure*} 

\section{Gyrokinetic code \texttt{stella} and coordinates}
\label{sec:stella}

The flux-tube $\delta\hspace{-0.1mm}f$-gyrokinetic code \texttt{stella} \cite{barnes2019stella} is employed, which has been extensively benchmarked for W7-X geometry \cite{gonzalez2022electrostatic}. Moreover, \texttt{stella} has been applied to the study of turbulent particle fluxes \cite{thienpondt2023prevention, alonso2023} and turbulent impurity transport \cite{garcia2021turbulent, garcia2021turbulent2, jose2023}. The code evolves electrostatic fluctuations by solving the nonlinear Vlasov and quasi-neutrality equations in five-dimensional phase space. We refer the reader to \cite{barnes2019stella, gonzalez2022electrostatic} for a detailed overview of the equations solved by \texttt{stella} and to \cite{gonzalez2022electrostatic} for the explanation of the notation and conventions that we follow here.
The magnetic field is written in Clebsch form \cite{d2012flux} as \mbox{$\mathbf{B} = (d\psi_t/dr) \, \boldsymbol{\nabla} \hspace{-0.2mm} r \times \boldsymbol{\nabla} \hspace{-0.2mm} \alpha $},  with \mbox{$r = a\sqrt{\psi_t/\psi_{t,\text{LCFS}}}$} a flux surface label, which is taken to be the effective minor radius coordinate (gray contours in \mbox{figure \ref{fig:geometry}}), and \mbox{$\alpha = \theta - \iota \zeta$}  a field line label (green lines in \mbox{figure \ref{fig:geometry}}). Here $\psi_t$ is the toroidal flux divided by $2\pi$,  $\psi_{t,\text{LCFS}}$ \mbox{is the value of $\psi_t$} at the last closed flux surface, $a$ is the minor radius,  $\iota(r)$ is the rotational transform, and $(\zeta,\theta)$ are straight-field line coordinates, which are chosen to be the cylindrical toroidal \mbox{angle $\zeta$} and the poloidal PEST \mbox{coordinate $\theta$ \cite{grimm1983ideal}}.   
The simulation domain is restricted to a flux tube (blue box in figure \ref{fig:geometry}), surrounding a chosen magnetic field line $\alpha_0$ (red curve in \mbox{figure \ref{fig:geometry}}). 
Since $(r,\alpha)$ are field-aligned coordinates, the cross-section of the flux tube is twisted into a parallelogram by the local magnetic shear as one moves along the field line. Note that the cross-section is taken to be rectangular at the center of the flux tube.

The phase-space coordinates employed by \texttt{stella} are denoted by $\{x,y,z,v_\parallel,\mu\}$, with $\{x,y,z\}$ a left-handed coordinate system. The local coordinates perpendicular to the magnetic field lines are defined as \mbox{$x=r - r_0$} and \mbox{$y=r_0(\alpha - \alpha_0)$}. The index `0' refers to quantities that are evaluated on the chosen magnetic field line. The parallel coordinate $z$ measures distances along the field line, hence $\theta$ or $\zeta$ can be used, and it is \mbox{normalized to $[-\pi,\pi]$}. The velocity coordinates consist of the parallel velocity $v_\parallel$ and the magnetic moment $\mu = v^2_\perp/2B$, with $v_\perp$ the perpendicular velocity and $B$ the magnetic field strength. 
The turbulent distribution function $g_s$ and the perturbed electrostatic \mbox{potential $\varphi$} are Fourier transformed with respect to the perpendicular $(x,y)$-coordinates.
The dimensions $(L_x,L_y)$ of the flux-tube cross-section define the wavenumbers that are included in the simulations, i.e., $k_x = 2\pi m/L_x$ and $k_y = 2\pi n/L_y$, where $m \in \mathbb{Z}$ and $n \in \mathbb{Z}$ are integers. Note that negative $k_y$ values are omitted as they can be recovered using the reality condition $\hat{\varphi}(k_x,k_y,z) = \hat{\varphi}^*(-k_x,-k_y,z)$, with $\hat{\varphi}_{\mathbf{k}}^*$ the complex conjugate of $\hat{\varphi}_{\mathbf{k}}$, and the hat denotes the Fourier coefficients of $\varphi$.

The perpendicular coordinates are normalized with respect to the gyroradius (or Larmor radius) \mbox{$\rho_i = v_{\text{th},i}/\Omega_i$} of the ion species, which are taken to be hydrogen nuclei in this work.  Here $v_{\text{th},i} = \sqrt{2T_i/m_i}$ is the reference thermal velocity, $\Omega_i = eB_\text{ref}/m_i$ is the reference gyrofrequency, \mbox{$B_\text{ref} = 2\,|\psi_{t,\text{LCFS}}|/a^2$} is the reference magnetic field, $T_i$ is the temperature  and $m_i$ is the mass of the hydrogen nuclei. The parallel coordinate is normalized with respect to the minor radius $a$ and the velocities are normalized with respect to the reference thermal velocity $v_{\text{th},i}$.    
The simulations presented in this paper are electrostatic, collisionless, consider ion-Larmor scales and include both kinetic hydrogen nuclei and electrons, with equal densities \mbox{($n_i=n_e$)} and \mbox{temperatures ($T_i=T_e$)}. Moreover, the ion and electron density gradients are assumed to be equal ($a/L_{n_i} =  a/L_{n_e}$) with $a/\hspace{-0.1mm}L_{n_s} = -(a/n_s)\hspace{0.2mm}dn_s/dr$ the normalized density gradient.

\section{Magnetic configurations}
\label{sec:devices}

To address the effect of the magnetic geometry on linear growth rates and nonlinear heat transport, this study incorporates the tokamak geometry of the Cyclone Base Case (CBC), which has been used extensively to benchmark gyrokinetic \mbox{codes \cite{dimits2000simulation}}, as well as the ASDEX Upgrade tokamak, constructed at IPP (Germany), which commenced operations \mbox{in 1991}.   
On the other hand, the \mbox{W7-X}, TJ-II, LHD and NCSX stellarators are included. \mbox{Wendelstein 7-X} is a low-shear five-field-period helias configuration, that has been in operation at IPP (Germany) since 2015, for which the standard EIM \mbox{configuration \cite{Geiger_2015}} is used. Another low-shear device is the four-field-period heliac configuration \mbox{TJ-II}, which has been in operation since 1997 at CIEMAT (Spain). The standard configuration (100\_44\_64) is considered. NIFS (Japan) houses the Large Helical Device,  a ten-field-period heliotron configuration, generating plasmas since 1998, for which the standard configuration with a major radius of \mbox{$R_0=3.72$} is used. Lastly, NCSX is a three-field-period quasi-symmetric stellarator, in particular featuring toroidal quasi-symmetry, also known as quasi-axisymmetry, developed at PPPL (United States). The magnetic configurations of AUG, TJ-II, LHD, NCSX and W7-X are generated by the three-dimensional MHD equilibium code \mbox{\texttt{VMEC} \cite{hirshman1983steepest}}, whereas the considered flux surface of CBC is defined by a set of Miller parameters \cite{miller1998noncircular}, explicitly given in \cite{barnes2019stella}. The configurations are taken to have $\beta = 0$, with $\beta$ the ratio of the plasma pressure to the magnetic pressure.

The simulations are performed at the radial location $r/a = 0.7$, for which the flux surfaces are shown in \mbox{figure \ref{fig:3D}}. The magnetic field strength is minimum (blue) on the outboard side and maximum (red) on the inboard side of the tokamaks. This trend does not necessarily hold in stellarators, where a minimum of the magnetic field can also be found on the inboard side, and vice versa.  An overview of the geometric characteristics of these flux surfaces is given in \mbox{table \ref{tab:geometry}}. 
For tokamaks, the strongest turbulence is expected to be found at the outboard midplane, therefore gyrokinetic studies generally focus on the $\alpha_0=0$ field line centered at \mbox{$(\theta,\zeta) = (0,0)$}, which lays on the bean-shaped cross-section of \mbox{W7-X} (see figure \ref{fig:geometry}), and the outboard midplane of the tokamaks. Note that the magnetic geometry and plasma parameters included in the simulations depend solely on their values and radial derivatives along the selected magnetic \mbox{field line $\alpha_0$}. Seeing that each field line in a stellarator has a unique magnetic geometry, different field lines will simulate different levels of \mbox{turbulence \cite{gonzalez2022electrostatic, sanchez2021gyrokinetic}}. Although in stellarators the turbulence may be strongest elsewhere on the flux surface, this study is limited to the $\alpha_0=0$ field line, since it is stellarator-symmetric \cite{dewar1998stellarator}, allowing us to utilize the generalized twist-and-shift boundary \mbox{conditions \cite{martin2018parallel}}, which are particularly advantageous for low-shear devices such as TJ-II and W7-X.
The flux tube is extended up to three poloidal turns for linear simulations, and up to approximately one poloidal turn for nonlinear simulations, where the exact length (see table \ref{tab:geometry}) is chosen to ensure that \mbox{$\Delta k_x = \Delta k_y$}. The resolutions used for the simulations are given in \ref{sec:resolution}, where the choices are justified  by performing rigorous convergence checks of the nonlinear heat fluxes, for the different devices and gradients considered in this work.

 \begin{table}[t!]
 	\caption{\label{tab:geometry} Geometric quantities at $r/a = 0.7$, with $a$ the minor radius, $R_0$ the major radius, $\hat{s} = (r/q)(dq/dr)$ the shear, $\iota=1/q$ the rotational transform and $q$ the safety factor. \mbox{$N_{f}$ is the} number of field periods of the device and $N_{\rm pol}$ is the length of the flux tube, given in number of poloidal turns, considered in the nonlinear simulations performed in section \ref{sec:nonlinear}.} \vspace{-3mm}
 	\lineup
 	\begin{indented}
 		\item[]\begin{tabular}{@{}lllllll}
 			\br
 			Device & $\0a$\,(m) & $\0R_0$\,(m)  & $\m\0\hat{s}$ & \m\0$\iota$ & $N_{f}$ & $N_{\rm pol}$\ \\
 			\mr
 			CBC   & \m 1.00 & \m 2.78 & \m 0.796 & \m 0.714 & \01 & 1.00 \\
 			AUG   & \m 0.63 & \m 1.62 & \m 1.587 & \m 0.744 & \01 & 1.00 \\
 			NCSX  & \m 0.32 & \m 1.44 & $-$0.703 & \m 0.541 & \03 & 0.95 \\
 			TJ-II & \m 0.19 & \m 1.50 & $-$0.057 & $-$1.592 & \04 & 1.19 \\
 			W7-X  & \m 0.51 & \m 5.51 & $-$0.127 & \m 0.894 & \05 & 1.11 \\
 			LHD   & \m 0.60 & \m 3.73 & $-$1.230 & $-$0.628 & 10  & 1.00 \\
 			\br  
 		\end{tabular}
 	\end{indented} 
 	\vspace{0.5mm}
 \end{table}

\begin{figure}[t!]
	\centering 
	\includegraphics[trim={0mm 1mm 0mm -1mm}] {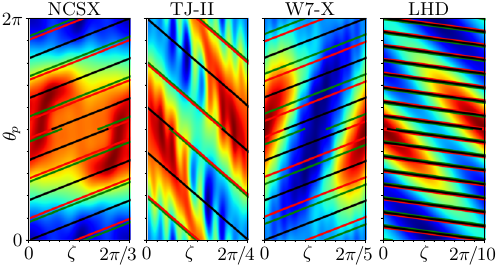} 
	\caption{Coverage of the flux surface $r/a=0.7$ by a flux tube that extends up to one (black), two (red) and three (green) poloidal turns.  Only one field period is shown for each device. \vspace{-2mm}} 
	\label{fig:coverage}
\end{figure} 

It is important to be aware of how densely the flux surface is covered by the simulated field line, which is shown in figure \ref{fig:coverage} for field lines extending up to one (black), two (red) and three (green) poloidal turns, centered at \mbox{$(\theta,\zeta) = (0,0)$}. The tokamaks are not shown, since they are axisymmetric, hence each poloidal turn is identical. Extending the field line for more than one poloidal turn samples new areas of the flux surface for \mbox{W7-X}. In contrast, for LHD, each additional turn overlaps almost exactly with the previous turn due to the specific value of the rotational transform on the $r/a=0.7$ flux surface. Therefore, extending the flux tube for multiple poloidal turns does not necessarily cover additional area of the flux surface. Nonetheless, it can be important to increase the length of the flux tube to capture modes that extend beyond the first poloidal turn of the device \cite{parisi2020toroidal}, or if the parallel correlation length is longer than the length covered by one poloidal turn \cite{beer1995field, ball2020eliminating}. \vspace{-1mm}

\begin{figure}[b]  
	\centering 
	\includegraphics[trim={0mm 2mm 0mm -2mm}]{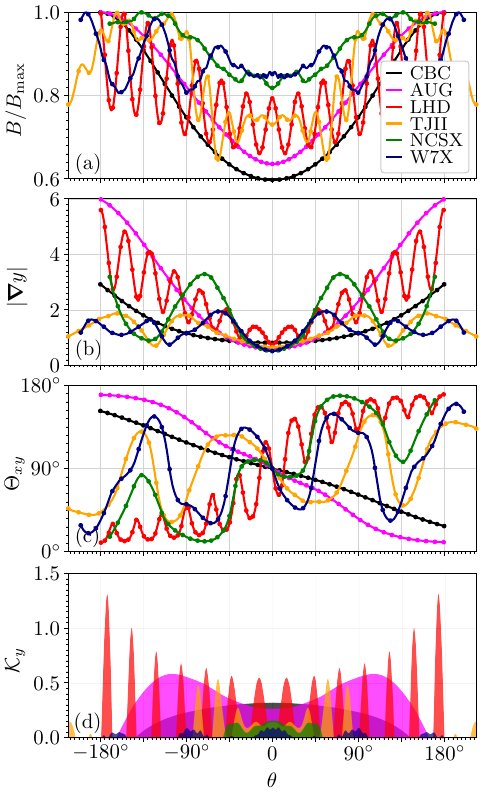}
	\caption{Geometric quantities along the chosen field line: (a) magnetic field strength normalized to its maximum along the field line, (b) magnitude of the perpendicular wavenumber for $k_x\rho_i=0.0$ and $k_y\rho_i=1.0$  (c) angle of the perpendicular cross-section $\Theta_{xy} = \text{arccos}(\boldsymbol{\nabla}x \cdot \boldsymbol{\nabla} y)/(|\boldsymbol{\nabla} x|\,|\boldsymbol{\nabla} y|)$ and (d) bad curvature regions for $k_x=0.0$ defined as $\mathcal{K}_y>0$ with \mbox{$\mathcal{K}_y = (a^2B_r/B^3) \,\mathbf{B} \times \boldsymbol{\nabla} B \cdot \boldsymbol{\nabla} y$}. The markers represent the parallel grid points used in the nonlinear simulations.} 
	\label{fig:fieldline} 
\end{figure}

\hfill
\section{Localization of micro-instabilities}
\label{sec:localization}

The background density and temperature gradients can drive a large number of micro-instabilities in fusion plasmas. Specifically, the ion-scale heat transport is predominately attributed to turbulence excited by ion-temperature-gradient-driven (ITG) and trapped-electron-mode (TEM) instabilities. These instabilities are destabilized or stabilized along the magnetic field lines depending on geometric characteristics such as magnetic wells ($B$), curvature and magnetic field inhomogeneity ($\mathcal{K}_y$) and shear (whose effects are seen in $|\boldsymbol{\nabla} y|$ and $\Theta_{xy}$). These characteristics, illustrated in figure \ref{fig:fieldline} for the field lines considered in the nonlinear simulations, are discussed in the following paragraphs.

\begin{figure}[h!] 
	\centering 
	\includegraphics[trim={0mm 1mm 0mm -3mm}]{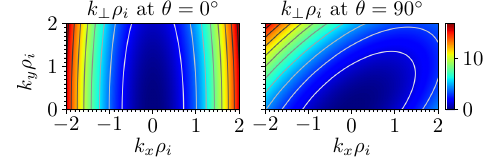} 
	\caption{Magnitude of the perpendicular wavenumber in \mbox{W7-X} for the $\alpha_0=0$ field line centered at  $(\theta,\zeta) = (0,0)$ at the parallel location $\theta=0^\circ$ (left) and $\theta = 90^\circ$ (right).} 
	\label{fig:kperp} 
\end{figure}

The magnetic shear, which gives a measure of how the rotational transform (or field line pitch) changes across neighboring flux surfaces, can have a stabilizing effect on the instabilities. Since field-aligned coordinates are employed, the cross-section of the flux tube is twisted into a parallelogram as one moves along the field line, as a result of the local magnetic shear, which is illustrated in figure \ref{fig:geometry}. The angle $\Theta_{xy}$ between $\boldsymbol{\nabla}x$ and $\boldsymbol{\nabla}y$ is depicted in \mbox{figure \ref{fig:fieldline}c}. In regions where the local magnetic shear is large, the cross-section, and thus the turbulent eddies, become very twisted and are subsequently suppressed through finite-Larmor-radius (FLR) effects. The FLR-effect is enforced via the zeroth order Bessel function of the first kind, $J_0(k_\perp\rho_s)$, appearing in the gyrokinetic equation as a result of averaging out the gyrophase \cite{catto1978linearized}. Here, \mbox{$\mathbf{k}_\perp = k_x \!\boldsymbol{\nabla} x + k_y\! \boldsymbol{\nabla}y$} is the perpendicular wavevector and $\smash{k_\perp =  ({k_x^2 \,|\boldsymbol{\nabla} x|^2 + 2\,k_x k_y \boldsymbol{\nabla} x \cdot \boldsymbol{\nabla} y + k_y^2 |\boldsymbol{\nabla} y |^2})^{1/2}}$ is its magnitude. The $J_0(k_\perp\rho_s)$-term damps small-scale fluctuations since $J_0(k_\perp\rho_s) \rightarrow 0$ for \mbox{$k_y\rho_i \gg 1$}, while large-scale fluctuations remain undamped. 
Therefore, an instability with wavenumbers $(\tilde{k}_x,\tilde{k}_y)$ is typically localized in regions along the field line where $k_\perp\rho_s(\tilde{k}_x,\tilde{k}_y,z)$ is minimum. Vice versa, if the strongest instability drives are found at a certain location $\tilde{z}$, then the most unstable mode, is most likely the $(k_x,k_y)$-mode that minimizes $k_\perp\rho_s(k_x,k_y,\tilde{z})$, which lies inside the inner contours shown in \mbox{figure \ref{fig:kperp}}. Note that the FLR-damping is much stronger for ions than for electrons since \mbox{$\rho_i = \sqrt{m_e/m_i}\,\rho_e$}.
Seeing that the center of the field line is typically chosen to have the largest instability drives, and the cross-section is taken to be rectangular in this location, the most unstable modes are generally found along $k_x = 0$ (see sections \ref{sec:gammavskxky} and \ref{sec:heat}), because \mbox{$k_x = 0$} minimizes $k_\perp\rho_s(k_x,{k}_y,z)$ if \mbox{$\boldsymbol{\nabla} \hspace{-0.2mm} x \cdot \boldsymbol{\nabla} \hspace{-0.2mm} y = 0$}. This is illustrated on the left side of \mbox{figure \ref{fig:kperp}}.
Moreover, the modes with $k_x = 0$ are confined to the center of the field line in high-shear devices. For these modes the perpendicular wavenumber reduces to $|k_y\boldsymbol{\nabla} \hspace{-0.2mm} y| \approx  | k_y (r_0\boldsymbol{\nabla} \hspace{-0.2mm}\theta - r_0 \iota \boldsymbol{\nabla} \hspace{-0.2mm}\zeta - \iota \hat{s}\zeta \boldsymbol{\nabla} \hspace{-0.2mm}x)|$ \mbox{(figure \ref{fig:fieldline}b)} which increases as one moves away from $\zeta=0$. As a result, the FLR-effect confines instabilities with \mbox{$k_x=0$} to the center of the field line if the global magnetic shear $\hat{s}$ (table \ref{tab:geometry}) is large.
On the other hand, modes with a finite radial wavenumber $k_x\neq 0$ typically have a minimum in $k_\perp\rho_s(\tilde{k}_x,\tilde{k}_y,z)$ further along the field line, hence these modes most likely peak away from $\zeta=0$. Therefore, it is important to include a wide range of radial wavenumbers in order to capture modes which are not excited at the center of the field line.

\begin{figure}[b]  
	\centering 
	\includegraphics[trim={0mm 1mm 0mm 2mm}] {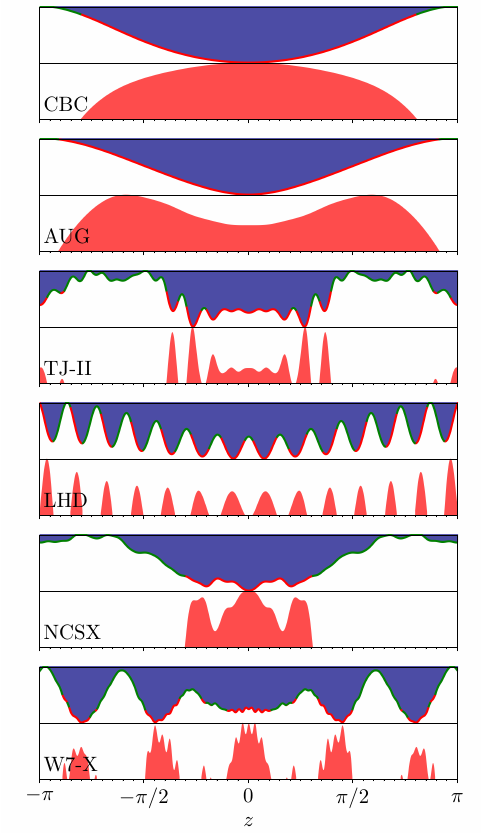} 
	\caption{Magnetic field strength (top rows) and bad curvature ($\mathcal{K}_y\!>\!0$) (bottom rows), where the $B$-line is colored red if it has bad curvature, and green if it has good curvature.}
	\label{fig:trappingcurvature}
\end{figure}

Besides the effects of the local and global magnetic shear, the magnetic wells, as well as the curvature and the inhomogeneity of the magnetic field, also play an important role on the localization of the instabilities. In particular, toroidal ITG \mbox{modes \cite{plunk2014collisionless}} are localized in regions of bad (destabilizing) curvature, for which \mbox{$k_y T'_i/\psi'_t(\mathbf{B} \times \boldsymbol{\nabla}\hspace{-0.3mm} B) \cdot \mathbf{k}_\perp > 0$} with \mbox{$T'_i = dT_i/dr <0$} since we only consider peaked temperature profiles and $\psi'_t=d\psi_t/dr<0$ in the configurations considered in this work. The most unstable modes typically have $k_x = 0$ (but not always, as we will see in section \ref{sec:gammavskxky}) for which the condition of bad curvature reduces to $(\mathbf{B} \times \boldsymbol{\nabla}\hspace{-0.3mm} B) \cdot \boldsymbol{\nabla} y > 0$. For simplicity, we define $\mathcal{K}_y = (a^2B_r/B^3) \,(\mathbf{B} \times \boldsymbol{\nabla}\hspace{-0.3mm} B) \cdot \boldsymbol{\nabla} y$, and the corresponding areas of bad curvature, where \mbox{$\mathcal{K}_y>0$}, are highlighted in figure \ref{fig:fieldline}d along the chosen magnetic field lines. Similar to ITG modes, which are destabilized by local bad curvature,  \mbox{trapped-electron-modes \cite{helander2013collisionless}} are driven unstable by the bounce-averaged bad curvature, which is the bad curvature that a particle trapped in a magnetic well experiences between two consecutive bounce points. Moreover, TEMs reside in regions where electrons are trapped, which is inside the magnetic wells shown in figure \ref{fig:fieldline}a.
Both the TEM and ITG mode can thus be stabilized by reducing the magnitude of the bad curvature and by reducing the connection length between regions of good and bad curvature. Moreover, the separation of regions of trapping and bad curvature has a stabilizing effect on TEMs.

Figure \ref{fig:trappingcurvature} shows the regions of particle trapping (top rows) and bad curvature (bottom rows) along the chosen magnetic field lines, for modes with $k_x=0$, in order to discuss their overlap. Specifically, for CBC, \mbox{TJ-II} and NCSX, the location with the deepest magnetic well has the most pronounced bad curvature. On the other hand, in AUG and LHD these locations do not coincide. Here, the most pronounced bad curvature is found further along the field line, for $\zeta \neq 0$, because of the $\iota \hat{s} \zeta \boldsymbol{\nabla} \hspace{-0.2mm}x$ term inside $\mathcal{K}_y$. Finally, in W7-X, the central well of the chosen field line corresponds to the most pronounced bad curvature, while the deepest wells are found at \mbox{$z = \pm 1.41$ (and $z = \pm 2.51$)} where the curvature is comparable to the center.  
Since bad curvature ($\mathcal{K}_y\!>\!0$) is destabilizing and good curvature ($\mathcal{K}_y\!<0$) is stabilizing, it is important to quantify the amount of bad curvature that occurs along the field line for ITGs, while the amount of bad curvature inside the magnetic wells is decisive for TEMs. It can be seen that the field lines of the tokamaks consist almost entirely of bad curvature, since  $\hat{s} = 0.8$ in CBC and $\hat{s} = 1.6$ in AUG, which increases the region of bad curvature through the $\iota \hat{s} \zeta \boldsymbol{\nabla} \hspace{-0.2mm}x$ term in $\mathcal{K}_y$, whereas the bad curvature region would be confined to the outboard side ($\theta \in [-90^\circ,90^\circ]$) if $\hat{s}=0$. 
In contrast, only 30-50\% of the field lines in the stellarators are covered by bad curvature. In particular, in NCSX and \mbox{W7-X} the magnetic wells have a considerable amount of good curvature, while in TJ-II the majority of its large magnetic well has bad curvature. Finally, in LHD the two inner magnetic wells consists almost entirely of bad curvature, whereas for the outer magnetic wells, the regions of bad curvature have been shifted with respect to the regions of trapping because of the large global magnetic shear ($\hat{s} = -1.2$).

\section{Linear stability analysis}
\label{sec:linear}		

Linear gyrokinetic simulations are performed for the six considered devices in order to discuss the micro-instabilities that are driven unstable in the presence of various density and temperature gradients.
To start, in section \ref{sec:gammavskxky}, scans along $(k_x,k_y)$ are performed considering $a/L_{T_i}=3.0$, $a/L_{T_e}=0.0$ and various density gradients to confirm that the most unstable mode is generally found for $k_x=0.0$. Next, detailed scans for $0.0 \leq a/L_n \leq 4.0$ are conducted to study the evolution of the growth rate with the density gradient in section \ref{sec:linearlineallkys}, considering \mbox{$k_x=0.0$} and limiting the binormal wavenumber to ion scales, $k_y\rho_i\leq 2.0$. The most unstable mode is identified and discussed in section \ref{sec:mostunstable}, for which the contributions of the ion temperature gradient and the density gradient are isolated in section \ref{sec:linearlineITGTEM}. A vanishing electron temperature gradient is considered in the simulations to avoid the excitation of the ETG mode, which is isomorphic to the ITG mode, hence it is driven at electron scales, \mbox{$k_{y,\text{ETG}} = (\rho_i/\rho_e)\,k_{y,\text{ITG}}$}, and it has much larger growth rates, $\gamma_\text{ETG} = (\rho_i/\rho_e) \gamma_\text{ITG}$. Including electron scales is beyond the scope of this work. Nonetheless, the effect of $a/L_{T_e}=3.0$ on the ion-scale instabilities is discussed in \mbox{section \ref{sec:linearlineITGTEM}}. 
Finally, in section \ref{sec:gammavsfprimtiprim} growth-rate maps are constructed for the six considered devices considering a wide range of gradients with $a/L_n \in [0.0, 8.0]$ and $a/L_{T_i} \in [0.0, 8.0]$.

\subsection{Growth rate as a function of $k_x\rho_i$ and $k_y\rho_i$}
\label{sec:gammavskxky}

\begin{figure}[b] 
	\centering 
	\includegraphics[trim={0mm 0mm 0mm 1mm}] {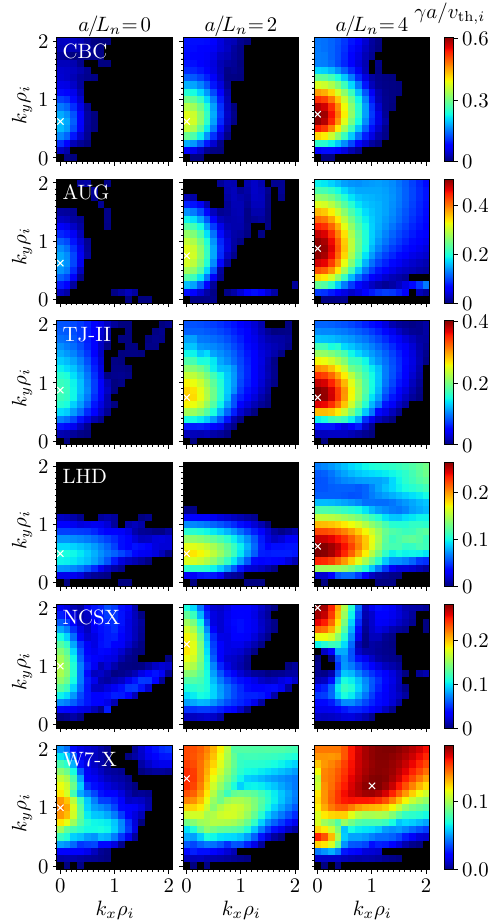} 
	\caption{Normalized growth rate as a function of $(k_x\rho_i, k_y\rho_i)$ for $a/L_{T_i}=3.0$, $a/L_{T_e}=0.0$ and $a/L_n=0.0$ (left column), $a/L_n = 2.0$ (middle column) and $a/L_n=4.0$ (right column). The most unstable mode is highlighted with a white cross. \vspace{0mm}}
	\label{fig:gamma_kxky}
\end{figure} 

Generally, linear studies of gyrokinetic simulations are performed considering a vanishing radial wavenumber, i.e., $k_x=0.0$. This simplification is justified when the most unstable mode peaks at a location along the flux tube where the flux-tube cross-section is perpendicular, or if it peaks symmetrically around such a location. Therefore, most gyrokinetic codes ensure that the center of the flux tube has a perpendicular cross-section, i.e., $\boldsymbol{\nabla} x \cdot \boldsymbol{\nabla} y = 0$ at $z=0$. Moreover, the center is typically chosen to correspond to the location with the deepest magnetic well, and the most pronounced bad curvature, as discussed in section \ref{sec:localization}. With these choices, the most unstable mode most likely peaks at the center, where it would be characterized by having $k_x = 0.0$, whereas if it would be localized further along the flux tube, at a location where the cross-section has become twisted, it is characterized by $k_x \neq 0.0$, due to the FLR-effects (see section \ref{sec:localization}).
In order to verify whether the most unstable modes have $k_x=0.0$, the growth rates $\gamma a/v_{\rm th,i}$ of the linear instabilities are shown in figure \ref{fig:gamma_kxky} as a function of $(k_x, k_y)$ for $a/L_{T_i}=3.0$, $a/L_{T_e}=0.0$ when considering a density gradient of $a/L_n=0.0$ (left column), $a/L_n=2.0$ (middle column) and $a/L_n=4.0$ (right column). The corresponding frequencies $\omega a/v_{\rm th,i}$ can be found in \ref{app:omega}. The perpendicular wavenumbers are restricted to ion scales, specifically, $0 \leq k_x\rho_i \leq 2.0$ and $0 \leq k_y\rho_i \leq 2.0$, where we limit the scans to positive $k_x$ since the field lines are stellarator-symmetric, hence the $k_x$-modes and the $-k_x$-modes have the same growth rates and frequencies, and we only consider positive $k_y$-modes because of the reality condition. The length of the flux tube is restricted to one poloidal turn in this section. 

For the three considered density gradients, the most unstable mode, which is highlighted by a white cross in figure \ref{fig:gamma_kxky}, is indeed found at $k_x=0.0$ in CBC, AUG, NCSX, LHD and TJ-II.
For \mbox{W7-X}, when considering low density gradients, the most unstable mode is also found at $k_x\rho_i = 0.0$. However, when the density gradient is increased to $a/L_n=4.0$, the fastest growing mode is instead found at $k_x\rho_i = 1.0$ and $k_y\rho_i = 1.375$.  Specifically, for $a/L_n=4.0$, the growth rate at \mbox{$k_x\rho_i=0.0$} and $k_y\rho_i=0.5$ is 13\% smaller than at $k_x\rho_i = 1.0$ and $k_y\rho_i=1.375$. As explained in \mbox{section \ref{sec:localization}}, the most unstable mode is most likely excited at the center for W7-X, but it can also be excited in the magnetic wells at $z = \pm 1.41$ (figure \ref{fig:trappingcurvature}) due to the depth of the magnetic well and the strength of the bad curvature at this location. At $z = \pm 1.41$ the cross-section has become sheared and the most unstable mode corresponds to the $(k_x,k_y)$-mode for which the FLR-damping is the smallest, which is inside the inner contour shown on the right-hand-side of figure \ref{fig:kperp}.

\begin{figure}[b!] 
	\centering 
	\includegraphics[trim={0mm 2.5mm 0mm 3.5mm}] {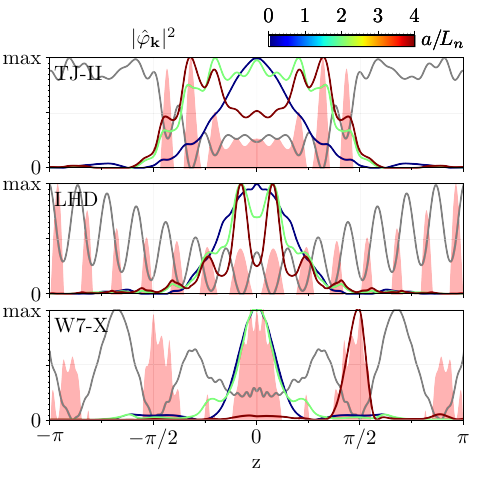} 
	\caption{Potential squared along the magnetic field line, for the most unstable mode (white crosses) of figure \ref{fig:gamma_kxky}. The strength of the magnetic field is shown in gray, and the regions of bad curvature are highlighted in red. \vspace{2.4mm}}
	\label{fig:linear_phi_vs_z}
\end{figure} 

In CBC, AUG and NCSX the most unstable mode peaks at the center of the flux tube where the cross-section is perpendicular, hence it has $k_x=0.0$. On the other hand, in TJ-II and LHD, for $a/L_n = \{2.0,4.0\}$, the mode with $k_x=0.0$ peaks symmetrically around the center of the flux tube, in the two magnetic wells at $z = \pm 0.23$ for LHD, and in the magnetic wells at $z = \pm 0.54$ and $z = \pm 0.83$ for TJ-II. This is illustrated in figure \ref{fig:linear_phi_vs_z} by plotting the parallel mode structure of the most unstable mode, corresponding to the white crosses in figure \ref{fig:gamma_kxky}. Finally, in W7-X, the most unstable mode has $k_x\rho_i \neq 0.0$ for $a/L_n=4.0$, for which the mode peaks in the magnetic well at $z =  1.41$. 

To conclude, if the cross-section at the center of the flux tube is chosen to be perpendicular and if it corresponds to the location with the strongest instability drives, then the most unstable mode is generally found for $k_x\rho_i=0.0$. However, in some cases, the most unstable mode is characterized by $k_x\rho_i \neq 0.0$, because the mode peaks further along the flux tube, at a location where the cross-section has become sheared.

\subsection{Growth rate as a function of $a/L_n$ and $k_y\rho_i$}
\label{sec:linearlineallkys}

\begin{figure}[!b]
	\centering 
	\includegraphics[trim={0mm 1mm 0mm 0mm}] {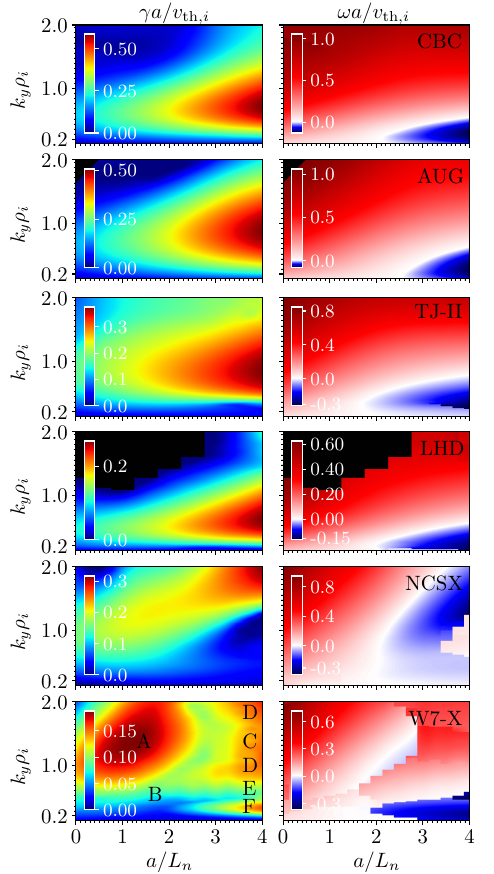} 
	\caption{Normalized growth rate (left) and frequency (right) as a function of the normalized density gradient, and the binormal wavenumber, considering $k_x\rho_i=0.0$,  $a/L_{T_i}=3.0$ and \mbox{$a/L_{T_e}=0.0$}. The frequency branches in W7-X are labeled A-F.  }
	\label{fig:gamma_map_line}
\end{figure}

The linear analysis is continued considering \mbox{$k_x\rho_i=0.0$}, while scanning $k_y\rho_i$ in $[0.125, 2.0]$ to investigate the perpendicular scale of the fluctuations. We simultaneously scan $a/L_n$ in $[0.0,4.0]$ to study the evolution of the instabilities as the density gradient is increased. An ion temperature gradient of \mbox{$a/L_{T_i}=3.0$} and a vanishing electron temperature gradient ($a/L_{T_e}=0.0$) are considered. The flux tube is extended up to 3 poloidal turns in order to capture extended modes. The growth rates and frequencies of the modes are shown in figure \ref{fig:gamma_map_line} for the six considered devices. Note that negative frequencies correspond to modes traveling in the electron diamagnetic direction, while those with positive frequencies travel in the ion diamagnetic direction.

\newpage
In CBC, AUG, TJ-II and LHD, the linear instabilities become increasingly more unstable with increasing density gradients. Moreover, CBC, AUG, TJ-II and LHD have continuous frequency maps, except for very small $k_y\rho_i$ in LHD and TJ-II. Note that the continuous change in the frequency is accompanied by a continuous change in the parallel mode structure of these modes, which is discussed in detail for the most unstable mode along $k_y\rho_i$ in the next section (figure \ref{fig:linear_phi_vs_z2}). Considering the range of scanned density gradients, it is likely that distinct instabilities are driven in these devices, such as the ITG mode \mbox{(at $a/L_n=0.0$)} and the TEM (at $a/L_n=4.0$), which are transforming into one another through a continuous change of the density \mbox{gradient \cite{kammerer2008exceptional}}.  

In contrast, the growth rate and frequency spectra of NCSX and W7-X are very different. Specifically, in NCSX, the growth rate also increases with increasing density gradients, however, the most unstable mode migrates to \mbox{larger $k_y\rho_i$}, in contrast to the other five devices. According to mixing length arguments, where \mbox{$Q_i \sim \gamma/k_y^2$}, one could expect the heat transport to be reduced as the density gradient increases  because the scale of the fluctuations becomes smaller. Moreover, note that a second instability with positive frequencies shows up at large density gradients.
On the other hand, the growth rate spectrum of W7-X consists of multiple different peaks, which correspond to six distinct frequency branches, characterized by having different parallel mode structures (shown for the most unstable mode along $k_y\rho_i$ in figure \ref{fig:linear_phi_vs_z2}). For the highest density gradient, $a/L_n=4.0$, four distinct instability branches co-exist within  \mbox{$k_y\rho_i \leq 2.0$}.  The different branches have been labeled (A) to (F).  In W7-X, the growth rate thus initially increases with the density gradient until $a/L_n = 1.25$, after which the growth rate decreases up to $a/L_n = 2.75$, due to the stabilization of instability (A) by increasing density gradients---which is most likely an ITG mode---without driving the other instabilities particularly unstable yet. As the density gradient is increased further, the instabilities corresponding to frequency branches (C) to (F) appear, which are driven unstable by the density gradient. Note that instability (F) has large perpendicular wavelengths, hence it could be particularly detrimental to the heat transport. 

In summary, CBC, AUG, TJ-II and LHD have  continuous frequency spectra, where the modes undergo a smooth transition from one instability into another as the density gradient is increased, whereas NCSX and W7-X are characterized by abrupt transitions between different frequency branches.  
Hence, it is clear that in most cases, a frequency branch does not necessarily correspond to a distinct analytical instability such as the ITG or TEM, and the identification of the instabilities should be done very carefully or be avoided altogether.

\subsection{Most unstable mode as a function of $a/L_n$}
\label{sec:mostunstable}

\begin{figure}[b]
	\centering 
	\includegraphics[trim={0mm 2mm 0mm 0mm}] {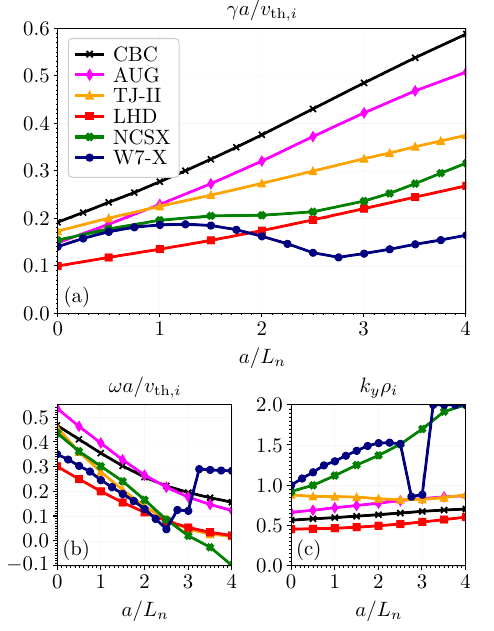} 
	\caption{The (a) normalized growth rate, (b) normalized frequency, and (c) normalized binormal wavenumber of the most unstable mode as a function of the normalized density gradient, $a/L_n$, considering $k_x = 0.0$, $a/L_{T_i}=3.0$ and  $a/L_{T_e}=0.0$.}
	\label{fig:gamma_vs_aLn}
\end{figure} 

Next, the growth rate, frequency and wavenumber of the most unstable modes of figure \ref{fig:gamma_map_line} are plotted as a function of the density gradient in figure \ref{fig:gamma_vs_aLn}, which is a linear analysis commonly performed in gyrokinetic studies. Recall that \mbox{we consider  $a/L_{T_i}=3.0$}, \mbox{$a/L_{T_e} = 0.0$, $k_x\rho_i=0.0$} and \mbox{$k_y\rho_i \leq 2.0$}. 
For CBC, AUG, TJ-II and LHD, the growth rate of the most unstable mode increases approximately linearly with the density gradient (figure \ref{fig:gamma_vs_aLn}a), while there is no significant change in the scale of the perpendicular fluctuations  (figure \ref{fig:gamma_vs_aLn}c). 
In NCSX, the growth rate also increases as the density gradient is increased, however, it has a weaker dependence \mbox{on $a/L_n$}. Additionally, the perpendicular fluctuations become smaller.  In contrast, in W7-X, the growth rates are reduced by 36\% from $a/L_n=1.25$ to $a/L_n=2.75$. A decrease in the perpendicular scales of the fluctuations is also observed in W7-X  in figure \ref{fig:gamma_vs_aLn}c, however, this is misleading since \mbox{figure \ref{fig:gamma_map_line}} indicates that multiple instabilities co-exist at high density gradients, some of which are characterized by having large perpendicular scales. Therefore, one has to be careful when interpreting linear results based only on the most unstable $(k_x,k_y)$-mode.
Nonetheless, the linear analysis suggest that a reduction of the heat transport with increasing density gradients is expected for NCSX and W7-X based on mixing length arguments, $Q_i \sim \gamma/k_y^2$, while an increase is expected for CBC, AUG, TJ-II and LHD.

\begin{figure}[b] 
	\centering 
	\includegraphics[trim={0mm 2mm 0mm 0mm}] {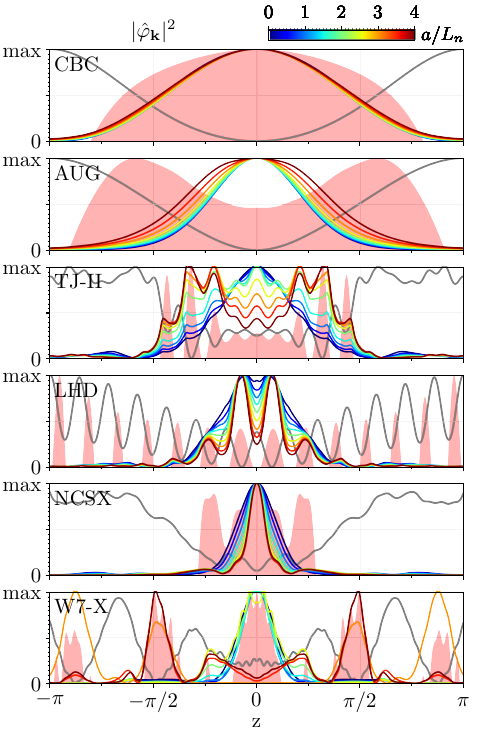} 
	\caption{Potential squared along the magnetic field line, for the modes in figure \ref{fig:gamma_vs_aLn}. The strength of the magnetic field is shown in gray, and the regions of bad curvature in red.}
	\label{fig:linear_phi_vs_z2}
\end{figure} 

In order to explain the differences in the linear behavior between these devices, $|\hat{\varphi}_\mathbf{k}|^2$ is plotted along the field line in figure \ref{fig:linear_phi_vs_z2}, for the modes from \mbox{figure \ref{fig:gamma_vs_aLn}}. These parallel mode structures give insight into the geometric characteristics that drive the instability. In every device, except for LHD, the modes peak at the center of the field line for small density gradients, where the ITG mode is believed to be dominant. This is mainly due to the FLR-damping, which scales with $k_\perp\rho_i$ for the ions, and has a minimum at the center of the flux tube for \mbox{$k_x=0.0$} \mbox{(figure \ref{fig:fieldline}b)}. On the other hand, as the density gradient is increased, the modes become localized inside the deepest magnetic wells, since TEMs are believed to become dominant. 
Specifically, CBC and AUG have a single magnetic well consisting of mostly bad curvature, hence the modes have the same parallel mode structure regardless of the density gradient. On the other hand, in LHD, the modes become more localized inside the magnetic wells as the density gradient is increased since the trapped electrons start driving the instabilities. 
Note that the modes are more localized (the mode structures are more narrow) in AUG ($\hat{s}=1.6$) and LHD \mbox{($\hat{s}=1.2$)} compared to CBC ($\hat{s}=0.8$), due to the higher global shear and thus stronger FLR-damping towards the ends of the flux tube. NCSX shows improved confinement of the modes, despite the moderate global shear ($\hat{s}=0.7$) which is likely due to the very strong local magnetic shear \mbox{(figure \ref{fig:fieldline}b)}. This limits the modes to a region which is even smaller than the already quite narrow region of bad curvature. In contrast, the mode structures in \mbox{TJ-II} are quite broad, since the shear is low and there is a large central well which consists almost entirely of bad curvature.
\mbox{Finally, in W7-X}, distinct mode structures occur, corresponding to distinct frequency branches (figure \ref{fig:gamma_vs_aLn}). For low density gradients, the modes peak at the center where  the FLR-damping is the weakest and the bad curvature is the largest \mbox{(figures \ref{fig:fieldline}b and \ref{fig:fieldline}d)}. Whereas for $a/L_n=3.0$ the mode peaks inside the wells at $z= \pm 2.51$, and for $a/L_n \geq 3.5$ the modes peak in the wells at $z= \pm 1.41$.

Moreover, the magnetic geometry is correlated to the magnitude of the growth rates of the instabilities. Specifically, in the devices where the mode structures are confined to narrow regions of the field line, such as in LHD, NCSX and W7-X, the growth rates are the smallest (figure \ref{fig:gamma_vs_aLn}). Additionally, the high local magnetic shear in NCSX, and the  intermittent regions of bad and good curvature in the central well of W7-X have a stabilizing effect.  TJ-II has the largest growth rates of the stellarators due to its low shear and broad magnetic well, which is covered almost entirely by bad curvature. Finally, the tokamaks have the highest growth rates since most of the field line is driving the modes unstable, where AUG performs slightly better, most likely due to its higher shear.

\subsection{Separate contributions to the growth rate from the ion temperature gradient and the density gradient}
\label{sec:linearlineITGTEM}

\begin{figure}[b!]
	\centering 
	\includegraphics[trim={0mm 0mm 0mm 0mm}] {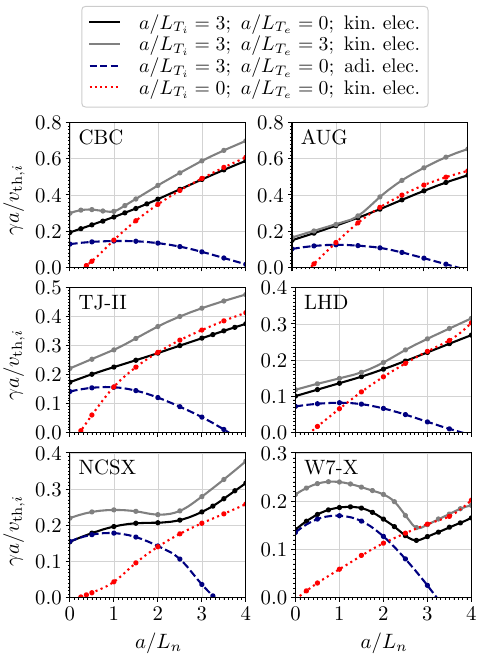} 
	\caption{Normalized growth rate as a function of the normalized density gradient with kinetic electrons, $a/L_{T_i}=3$ and a vanishing electron temperature gradient (black),  and with $a/L_{T_e}=3$ (gray). Ion-temperature-gradient-driven modes are isolated by treating the electrons as adiabatic (blue) and density-gradient-driven modes are isolated by setting $a/L_{T_i}=0$ (red).}
	\vspace{-3mm}
	\label{fig:linear_teprim}
\end{figure}

In most devices, we are unable to distinguish between different instability types, since the transitions between them are smooth (section \ref{sec:linearlineallkys}). Nonetheless, it is possible to isolate specific instabilities based on their instability drive by modifying the simulations accordingly, which is shown in figure \ref{fig:linear_teprim}. The original simulations considered in figure \ref{fig:gamma_vs_aLn} are reproduced here with black lines. Firstly, we consider the case of adiabatic electrons to isolate the instabilities which are driven by the ion temperature gradient and which rely solely on the ion dynamics (blue dashed lines).  Next, we impose vanishing temperature gradients while treating the electrons kinetically, to isolate modes which are driven by the density gradient and require electron dynamics (red dotted lines).

\begin{figure}[!b]
	\centering  
	\includegraphics[trim={0mm 3mm 0mm 1mm}] {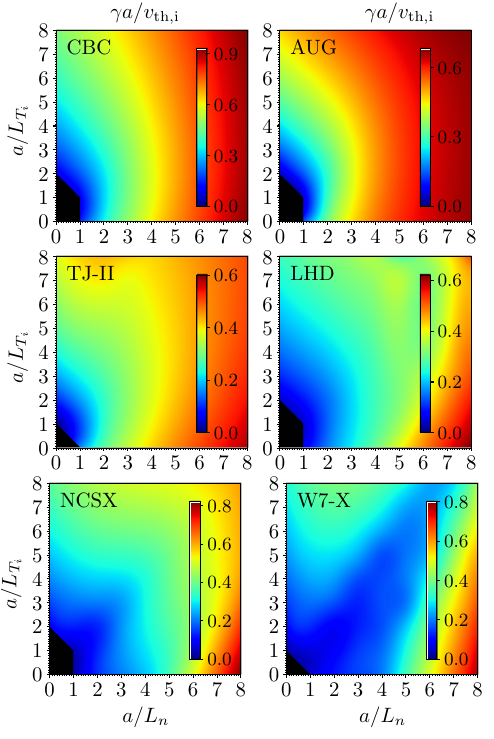} 
	\caption{\mbox{Normalized growth rate of the most unstable} mode for $k_x\rho_i=0.0$ and $k_y\rho_i \leq 2.0$, as a function of the \mbox{normalized density gradient} and the normalized ion temperature gradient, for a vanishing electron temperature gradient. }
	\label{fig:gamma_maps}
\end{figure}

\begin{figure}[b!]
	\centering    
	\includegraphics[trim={0mm 2mm 0mm 1mm}] {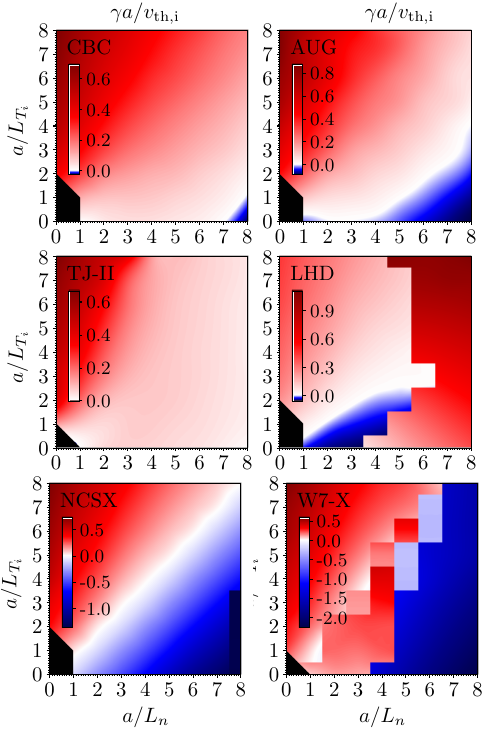}   
	\caption{Normalized frequency of the most unstable mode for $k_x\rho_i=0.0$ and $k_y\rho_i \leq 2.0$, as a function of the normalized density gradient and the normalized ion temperature gradient, for a vanishing electron temperature gradient.}
	\label{fig:omega_maps}
\end{figure}

The instabilities driven by the ions (blue dashed lines) show an initial increase in their growth rates with the density gradient until \mbox{$a/L_n = 1.0$}, after which the growth rates decrease, and eventually the instability is suppressed by the density gradient. The plasma parameters and behavior of these modes suggest that this instability corresponds to the toroidal ITG \mbox{mode \cite{plunk2014collisionless}}.  The density gradient for which these modes are suppressed is the highest in CBC, AUG, TJ-II and LHD, and the lowest in NCSX and \mbox{W7-X}.  Moreover, note that for $a/L_n=0.0$, including the electron dynamics \mbox{(blue lines $\rightarrow$ black lines)} has a destabilizing effect in every device, except in NCSX and W7-X, where it has no effect on the growth rate.  

The red dotted line in figure \ref{fig:linear_teprim} represents the contribution of the density-gradient-driven modes, which are most likely trapped-electron-modes (TEM) \cite{helander2013collisionless} but they could also be passing-particle-driven universal instabilities \cite{costello2023universal}, however this instability is typically subdominant to TEMs. A minimum density gradient is required in order to drive these modes unstable, after which the growth rate increases with the density gradient. Note that for high density gradients in W7-X, TJ-II and AUG, the linear growth rates are reduced by the presence of an ion temperature gradient (red lines $\rightarrow$ black lines), in contrast to NCSX, where the ion temperature gradient has a destabilizing effect. 

In CBC, AUG, TJ-II and LHD the growth rates of the simulations that include both the ion temperature gradient and kinetic electrons (black lines) increase monotonously with the density gradient. In contrast, in \mbox{W7-X}, and to lesser extent in NCSX, distinct instabilities are observed in the frequency spectra (figure \ref{fig:gamma_map_line}), and the growth rates of the original simulations (black lines) resembles the sum of the growth rates of the ion-temperature-gradient-driven modes (blue lines) and the density-gradient-driven modes \mbox{(red lines)}. 
The presence of a clear separation between these two contributions is related to the density gradient above which the growth rates driven solely by the density-gradient-driven modes exceeds those driven by the ion-temperature-gradient-driven modes, i.e., where the red line crosses the blue line. This occurs at around $a/L_n \approx 1.0$ for CBC, AUG,  TJ-II and LHD, whereas it occurs around around $a/L_n \approx 2.0$ for NCSX and W7-X. Hence, in NCSX and W7-X, there is a clear separation between these two instabilities, and there is a region of density gradients for which the  ion-temperature-gradient-driven modes are stabilized by the density gradient, without driving the density-gradient-driven modes particularly unstable yet.

Finally, the effect of adding a non-vanishing electron temperature gradient, set to \mbox{$a/L_{T_e} = a/L_{T_i} = 3.0$}, is represented by the gray lines in figure \ref{fig:linear_teprim}. It can be seen that the presence of an electron temperature gradient increases the growth rates in every device, hence it has a destabilizing effect on the ion-scale instabilities. Note that in CBC the ETG mode is dominant for $a/L_n \leq 0.75$, since the most unstable mode  is found at the end of the scanned wavenumber range, at $k_y\rho_i = 2.0$. Increasing the scanned $k_y\rho_i$-values shows that this growth rate belongs to the ETG branch which peaks at electron scales (not shown here).


\subsection{Growth rate as a function of $a/L_n$ and $a/L_{T_i}$}
\label{sec:gammavsfprimtiprim}

The linear analysis is extended to a wide range of plasma parameters, varying the gradients from flat density and ion temperature profiles ($a/L_n=0.0$ and $a/L_{T_i}=0.0$) up to extremely peaked profiles ($a/L_n=8.0$ and $a/L_{T_i}=8.0$), while considering a vanishing electron temperature gradient \mbox{($a/L_{T_e}=0.0$)}. The growth rates and frequencies of the most unstable modes, found for $k_x\rho_i=0.0$ and $k_y\rho_i \leq 2.0$, are shown in figures \ref{fig:gamma_maps} and \ref{fig:omega_maps}, respectively. The wavenumber of the most unstable mode is discussed in \ref{app:ky}. Note that the black regions indicate that no unstable modes were found within the considered CPU time. Resolving these modes, which lie near marginality, would require a more extensive study.  Moreover, it is important to acknowledge that for most devices, the gradients that are considered in this section are far beyond the experimentally relevant values at \mbox{$r/a=0.7$}. Therefore, one needs to be careful when drawing conclusions based on such extreme scans. 

It is clear that \mbox{W7-X} is the only device which exhibits a prominent \mbox{valley \cite{alcuson2020suppression}}, i.e., a region along the diagonal \mbox{$a/L_n \approx a/L_{T_i}$} which is characterized by having significantly reduced growth rates \mbox{(figure \ref{fig:gamma_maps})}. However, note that for the specific choice of \mbox{$a/L_{T_i}=3.0$} and $a/L_n \leq 4.0$, the reduction of the growth rate in W7-X is very modest \mbox{(figure \ref{fig:gamma_vs_aLn})}. As explained in the previous section, this valley in W7-X is formed by the stabilization of the ion-temperature-gradient-driven modes by increasing density gradients, without driving the density-gradient-driven modes particularly unstable yet.  
Similarly, a small reduction of the growth rate is observed along \mbox{$a/L_n \approx a/L_{T_i}$} in NCSX, while for the other four devices, no reduced growth rates are observed along the diagonal.

In figure \ref{fig:omega_maps}, the frequency of the most unstable mode is depicted. Each frequency branch, whose modes have a similar parallel mode structure and occur at a similar binormal wavenumber (see \mbox{\ref{app:ky}}), has been interpolated separately. In CBC, AUG, \mbox{TJ-II}, and NCSX the most unstable mode changes continuously from a pure ion-temperature-gradient-driven mode (top left) to a pure density-gradient-driven mode (bottom right), as can be seen by the continuous change of the frequency throughout all scanned gradients---except for the jump to a new frequency branch at \mbox{$a/L_n=8.0$} for NCSX. On the other hand, LHD exhibits a similar frequency map as the other devices for low density gradients, while for high density gradients, the most unstable mode belongs to a different frequency branch, which is characterized by a different parallel mode structure (see \mbox{\ref{app:ky}}).
In contrast, many distinct instabilities live inside the growth-rate valley of \mbox{W7-X}.

In summary, the linear analysis at $a/L_{T_i}=3.0$ reveals a modest reduction of the growth rates with increasing density gradients in \mbox{W7-X} and the absence of large-scale fluctuations at high density gradients in NCSX \mbox{(figure \ref{fig:gamma_map_line})}, indicating a potential reduction of nonlinear heat transport with increased density gradients.   In contrast, in CBC, AUG, TJ-II and LHD the linear observations point towards an increase of the heat transport with increasing density gradients, since the growth rates increase with $a/L_n$, and the perpendicular scale of the turbulence remains large. 
Moreover, analyzing the geometric characteristics of the chosen field line, it is evident that the modes in NCSX and \mbox{W7-X} are confined to narrow regions along the field lines, due to the high local magnetic shear in NCSX and due to the narrow magnetic wells and regions of bad curvature in W7-X (figure \ref{fig:linear_phi_vs_z2}).  		 

\section{Nonlinear analysis}
\label{sec:nonlinear}

In the final part of this paper, nonlinear simulations are performed to investigate the effect of density gradients on turbulence. Section \ref{sec:heat} provides a detailed discussion on the ion heat transport, while section \ref{sec:nonlinearITGTEM} isolates the contributions from the ion-temperature-gradient-driven modes and the density-gradient-driven modes. Next, section \ref{sec:nonlinearWithALTe} examines the electron heat flux, addressing the influence of a non-vanishing electron temperature gradient. Finally, more insight is given into the characteristics of the instabilities that drive the turbulence in section \ref{sec:distributon}, by identifying the contribution of the trapped and passing particles to the turbulent distribution function.  

\subsection{Turbulent heat transport}
\label{sec:heat}

\begin{figure}[!b]
	\centering 
	\includegraphics[trim={0mm 2mm 0mm 1mm}] {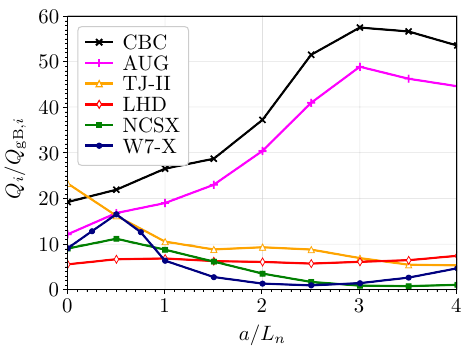} 
	\caption{Ion heat flux as a function of the normalized density gradient, considering $a/L_{T_i}=3.0$ and $a/L_{T_e}=0.0$.}
	\label{fig:Qi_vs_aLn}
\end{figure} 

\begin{figure}[b!]
	\centering
	\includegraphics[trim={0mm 1mm 0mm 1mm}] {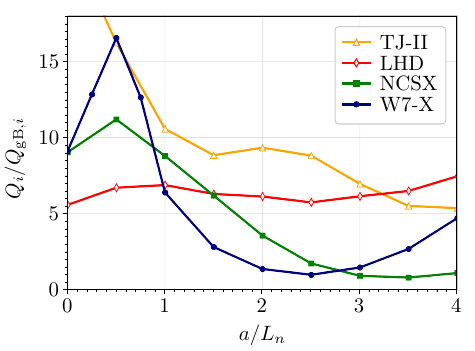} 
	\caption{Ion heat flux as a function of the normalized density gradient, for $a/L_{T_i}=3.0$ and $a/L_{T_e}=0.0$ (zoom of figure \ref{fig:Qi_vs_aLn}). \vspace{-1mm}}
	\label{fig:Qi_vs_aLn_zoom}
\end{figure}

In the following sections, we will discuss the ion and electron heat fluxes, $Q_i$ and $Q_e$, which are normalized with respect to the ion gyro-Bohm flux, defined as \mbox{$Q_{\rm gB,i}= n_i\,T_i\,v_{{\rm th},i}\,(\rho_i/a)^2$}. The precise definition of the heat flux in the context of flux-tube simulations can be found in \cite{gonzalez2022electrostatic}. The evolution of the ion heat flux as a function of the normalized density gradient, considering $a/L_{T_i}=3.0$ and $a/L_{T_e}=0.0$, is shown in figure \ref{fig:Qi_vs_aLn}. 
For the tokamak configurations, the density gradient gives rise to a significant increase of the ion heat transport, up to \mbox{$a/L_n = 3.0$}, after which it decreases slightly.  
In order to discuss the stellarators, a zoom of the this plot is shown in \mbox{figure \ref{fig:Qi_vs_aLn_zoom}}. For LHD, the ion heat flux is approximately independent of the density gradient. On the other hand, in TJ-II the ion heat flux is reduced significant for $a/L_n \leq 1.5$ and \mbox{$a/L_n\geq2.5$}, whereas it remains approximately constant within \mbox{$1.5 \leq a/L_n \leq 2.5$}. 
In contrast, in NCSX and W7-X, a very strong reduction of the ion heat flux is observed across a wide range of density gradients. Initially, the ion heat flux increases for $a/L_n\leq 0.5$, after which there is a reduction of the ion heat flux by an order of magnitude in both devices. Note that the ion heat flux starts increasing again for $a/L_n\geq 2.5$ in W7-X  and for $a/L_n \geq 3.5$ in NCSX. 

Comparing the nonlinear ion heat fluxes with the linear growth rates in figure \ref{fig:gamma_vs_aLn}a, it is clear that the reduction of the ion heat flux in W7-X is an order of magnitude stronger than the reduction of the growth rates. Moreover, the ion heat flux in NCSX is reduced very strongly, while the growth rates do not indicate any reduction. For LHD and \mbox{TJ-II} the growth rates increase linearly with the density gradient, while their ion heat flux is either constant or gets reduced. Therefore, it is clear that there is no correlation between the nonlinear ion heat fluxes and the linear growth rates for the stellarators. Even when taking into account the scale of the perpendicular fluctuations observed in the linear simulations (figure \ref{fig:gamma_vs_aLn}c), the reduction of the ion heat flux in NCSX and \mbox{W7-X} is much larger than the reduction of $\gamma/k_y^2$. Nonetheless, NCSX and W7-X are the only two stellarators for which the linear analysis pointed towards a reduction of the nonlinear heat transport, albeit the magnitude of the reduction was significantly underestimated. On the other hand, for the tokamaks, the increase of the ion heat flux is in agreement with the increase of the growth rates for $a/L_n \leq 3.0$.

\begin{figure}[h!]
	\centering
	\includegraphics[trim={0mm 3mm 0mm -1mm}] {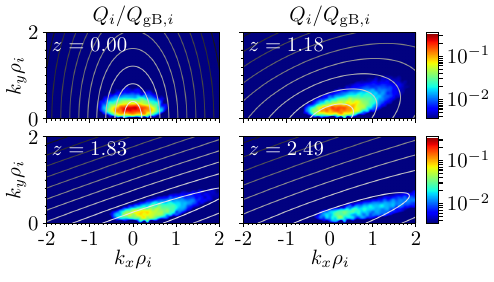}
	\caption{Contributions of the individual $(k_x,k_y)$-modes to the ion heat flux of LHD, for different points along the field line considering $a/L_n=4.0$, $a/L_{T_i}=3.0$ and $a/L_{T_e}=0.0$.  The contour lines represent the magnitude of the perpendicular wavenumber $k_\perp\rho_i$ (white is minimum, black is maximum).}
	\label{fig:kxkyspectraLHD}
\end{figure}

\begin{figure}[b!]
	\centering
	\includegraphics[trim={0mm 2mm 0mm 1mm}] {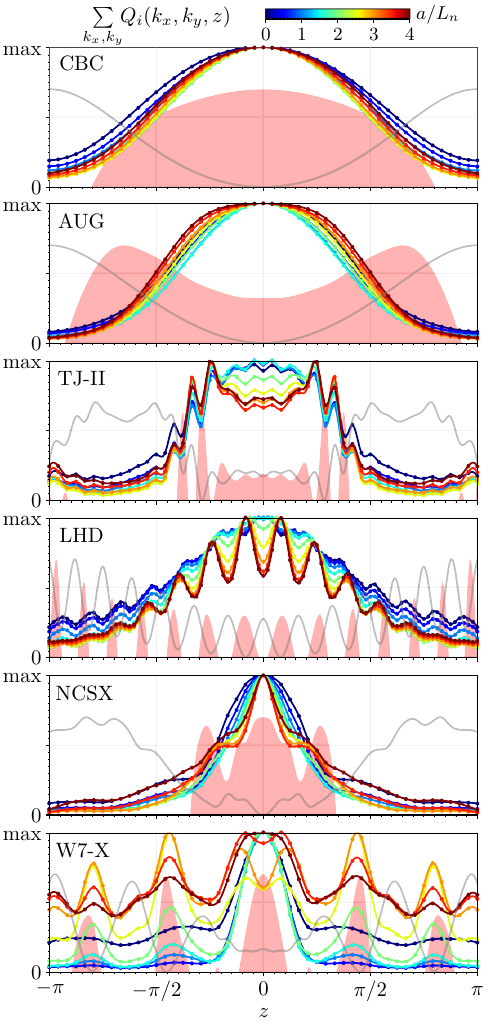} 
	\caption{Contributions of the individual $z$-points to the ion heat flux. The magnetic field strength (gray) and the regions of bad curvature for $k_x=0.0$ (red) are highlighted. }
	\label{fig:qfluxvsz}
\end{figure}

The contributions of the specific $(k_x,k_y)$-modes to the ion heat flux of LHD are shown in figure \ref{fig:kxkyspectraLHD} for four different locations along the field line, in order to illustrate the effect of the local magnetic shear and the FLR-damping. At the center of the field line $(z=0)$ the cross-section of the flux tube is perpendicular and the minimum perpendicular wavenumber $k_\perp\rho_i$ occurs for $k_x=0.0$ (see section \ref{sec:localization}). As one moves along the field line, the cross-section becomes sheared (figure \ref{fig:fieldline}c), and the minimum perpendicular wavenumber $k_\perp\rho_i$ is found along a diagonal in the $(k_x,k_y)$-plane. Figure \ref{fig:kxkyspectraLHD} shows that the modes that contribute the most to the heat flux are those for which the perpendicular wavenumber $k_\perp\rho_i$ is minimum---corresponding to the modes within the inner contour line---and the FLR-damping is thus the weakest. Moreover, as the local magnetic shear increases, the turbulent eddies are sheared away and the turbulence is reduced. Therefore, in most simulations, the heat flux peaks at the center of the field line, and the largest contributions arise from the $k_x=0.0$ modes. For this reason, the bad curvature regions are shown for $k_x=0.0$ in \mbox{figure \ref{fig:qfluxvsz}}. Note that other $k_x$-modes also contribute to the heat flux, for which the bad curvature regions are located elsewhere.

\begin{figure}[h!]
	\centering
	\includegraphics[trim={0mm 3mm 0mm 0mm}] {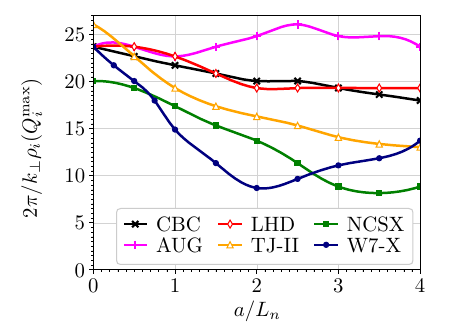}
	\caption{Dominant perpendicular scale of the fluctuations that contribute to the ion heat flux, for $a/L_{T_i}=3.0$ and $a/L_{T_e}=0.0$.} 
	\label{fig:dominantkperp}
\end{figure}

The contributions $Q_i(k_x,k_y,z)$ to the ion heat flux are transformed to $Q_i(k_\perp\rho_i)$, and the $k_\perp\rho_i$ for which $Q_i(k_\perp\rho_i)$ is maximum is considered the dominant scale of the perpendicular fluctuations of the turbulence, which is shown in figure \ref{fig:dominantkperp} as $2\pi/k_\perp\rho_i(Q_i^\textrm{max})$. \mbox{For W7-X}, NCSX, and to lesser extent for TJ-II, the perpendicular fluctuations become significantly smaller as the density gradient increases. Moreover, the trend of $2\pi/k_\perp\rho_i$ for W7-X, NCSX and TJ-II is very similar to the trend of their ion heat fluxes (figure \ref{fig:Qi_vs_aLn_zoom}), hence the heat losses are reduced partly because the scale of the turbulence becomes smaller. In contrast, for CBC and AUG the scale of the turbulence is not correlated to the trend of the ion heat flux, while in LHD both the ion heat flux and the scale of the fluctuations are approximately independent of the density gradient.

The contribution to the ion heat flux coming from specific points along the field line is visualized in figure \ref{fig:qfluxvsz}. In every device, except for W7-X, the localization of the ion heat losses does not change significantly when increasing the density gradient, besides the increased contribution coming from the magnetic wells. 
Specifically, for \mbox{$a/L_n=0.0$}, the ion heat losses are the largest at the center of the field line ($z=0$). As explained in section \ref{sec:localization}, the ITG mode, which is believed to be dominant for \mbox{$a/L_n=0.0$}, drives turbulence in locations which have the most pronounced bad curvature and the least FLR-damping, which typically corresponds to the center of the field line (figure \ref{fig:fieldline}).  For CBC, AUG and NCSX the center also corresponds to the deepest magnetic well, hence the heat losses remain dominant at $z=0$ as the density gradient is increased, and trapped-electron-modes are believed to become dominant. Whereas for W7-X, LHD and TJ-II the heat losses at high density gradients predominantly arise from locations further along the field line, corresponding to the deepest magnetic wells. 
Moreover, for W7-X the distribution of the ion heat losses along the field line changes multiple times as the density gradient is increased. For $a/L_n = 0.0$, the heat losses arise predominately from the center of the field line, however, a substantial portion of the heat losses also arise uniformly from the  entire field line. When a finite density gradient is introduced \mbox{$(a/L_n = \{0.5,1.0,1.5\})$},  the uniform contribution disappears and the heat flux is generated almost entirely at the center. For \mbox{$a/L_n=2.0$}, which is close to the minimum ion heat flux (figure \ref{fig:Qi_vs_aLn_zoom}), heat losses start arising from the trapping regions, indicating that trapped-electron-modes are beginning to become unstable. As the density gradient is increased further $(a/L_n = \{2.5,3.0\})$, heat is lost inside each magnetic well, however, the heat losses in the central well are localized predominantly on the sides of the magnetic well. This could be explained by the fact that trapped electrons spend most of the time in their orbits near their bouncing points, hence the heat losses are localized here. For the highest density gradients that are considered $(a/L_n = \{3.5,4.0\})$, the preference of the magnetic wells is diminished and instead the majority of heat losses arise uniformly from the entire field line, suggesting a significant involvement of passing electrons.

\begin{figure}[h!]
	\centering
	\includegraphics[trim={0mm 3mm 0mm 1mm}] {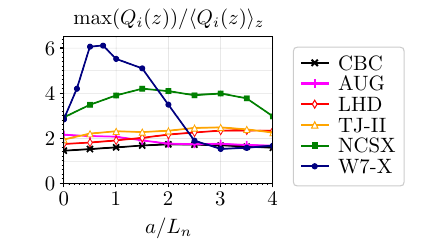}
	\caption{Ratio of the maximum ion heat flux along the field line with respect to the field-line averaged ion heat flux, as a function of the normalized density gradient, $a/L_n$. \vspace{-2mm}}
	\label{fig:Qi_vs_aLn_localization} 
\end{figure}

\newpage
To quantify how localized the heat losses are, the maximum ion heat flux along the field line is compared to the field-line averaged heat flux in figure \ref{fig:Qi_vs_aLn_localization}. For CBC, AUG, TJ-II and LHD, the maximum heat flux along the field line is around twice as big as the field-line averaged heat flux, regardless of the density gradient, because the heat losses remain localized at the center of the field line. Similarly, in NCSX the localization does not change with the density gradient, however, the heat losses are much more localized. This is likely due to the fact that the bad curvature region, as well as the deepest magnetic well, are rather narrow, and the strong local magnetic shear \mbox{(figure \ref{fig:fieldline}c)} damps the fluctuations arising from the two outer peaks of bad curvature. On the other hand, in \mbox{W7-X}, there is a uniform loss of heat along the entire field line, as well as a strong contribution at $z=0$ for $a/L_n=0.0$, after which the heat losses become strongly localized to the center of the field line as the density gradient is increased. The heat losses remain strongly localized until $a/L_n = 1.5$ after which a uniform contribution along the entire field line emerges, indicating that an instability driven by passing electrons is likely becoming unstable for $a/L_n \geq 2.0$.

\subsection{Separate contributions to $Q_s$ from the ion temperature gradient and the density gradient}
\label{sec:nonlinearITGTEM}

\begin{figure}[b!]
	\centering 
	\includegraphics[trim={0mm 1mm 0mm 0mm}] {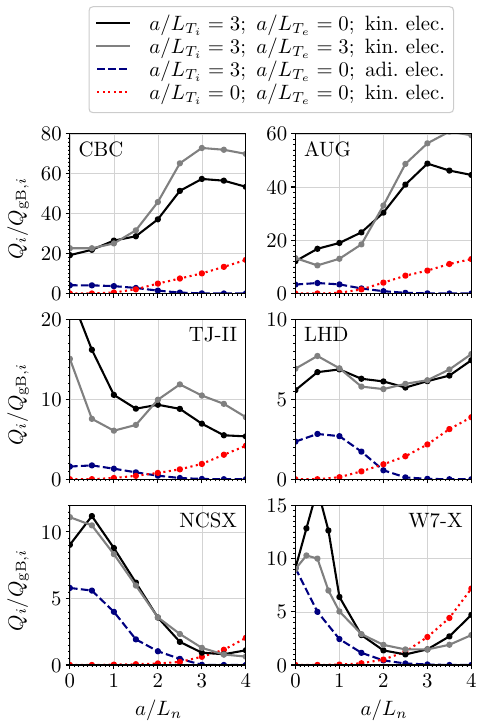}  
	\caption{Normalized ion heat flux as a function of the normalized density gradient with kinetic electrons for \mbox{$a/L_{T_e}=0.0$} (black) and  $a/L_{T_e}=3.0$ (gray). Ion-temperature-gradient driven modes are isolated by treating the electrons as adiabatic (blue) and density-gradient-driven modes are isolated by setting $a/L_{T_i} = 0.0$ (red), both considering \mbox{$a/L_{T_e}=0.0$}.}
	\label{fig:qi_contributions}
\end{figure}

\begin{figure}[b!]
	\centering
	\includegraphics[trim={0mm 0mm 0mm 16mm},clip] {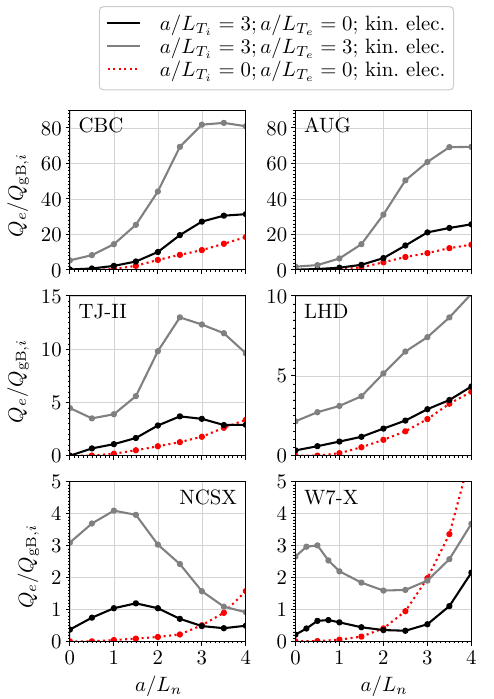} 
	\caption{Normalized electron heat flux as a \mbox{function of the} normalized density gradient for \mbox{$a/L_{T_e}=0.0$} (black) and  $a/L_{T_e}=3.0$ (gray). The density-gradient-driven modes are isolated by setting $a/L_{T_i} = a/L_{T_e} = 0.0$ (red). }
	\label{fig:qe_contributions}
\end{figure}

To gain more insights into the trends of the heat fluxes, the contributions of the ion-temperature-gradient-driven modes (blue dashed lines) and the density-gradient-driven modes (red dotted lines) to the ion and electron heat flux are isolated and shown in figures \ref{fig:qi_contributions} and \ref{fig:qe_contributions}, respectively. The simulations containing both the ion temperature gradient and kinetic electrons, shown in figure \ref{fig:Qi_vs_aLn}, are represented here by the black lines. The ion-temperature-gradient-driven modes are isolated by treating the electrons adiabatically, whereas the density-gradient-driven modes are isolated by considering vanishing temperature gradients. 

The ion-temperature-gradient-driven modes (blue dashed lines) are stabilized with increasing density gradients, in the same way as in the linear simulations (figure \ref{fig:linear_teprim}). Under these conditions the toroidal ITG mode is believed to be dominant. Note that the largest ion heat flux is driven in W7-X and NCSX at \mbox{$a/L_n=0.0$} and the lowest in LHD and TJ-II. 
In \mbox{TJ-II}, including kinetic electrons in the simulations (blue lines $\rightarrow$ black lines) drastically increases the ion heat flux at $a/L_n=0.0$, while a significant increase in the ion heat flux is observed in CBC and AUG, a moderate increase in LHD and NCSX and finally in W7-X the electron dynamics do not effect the ion heat flux at $a/L_n=0.0$. 

The ion heat flux driven by the density-gradient-driven modes (red dotted lines) increases with increasing density gradients, which is in line with the linear results (figure \ref{fig:linear_teprim}). In this case, the turbulence is believed to be driven primarily by trapped-electron-modes. These modes drive the lowest ion heat fluxes in  NCSX, and the highest in CBC and AUG. Moreover, in CBC and AUG the ion heat flux is increased significantly by the ion temperature gradient at $a/L_n=4.0$ (red line $\rightarrow$ black line), whereas in LHD it is increased moderately. Finally, in TJ-II, NCSX and W7-X, the effect of the ion temperature gradient is very small, moreover, it is capable of reducing the ion heat flux driven by the density-gradient-driven modes.

When including both kinetic electrons and the ion temperature gradient at the same time (black lines), the ion heat flux can resemble a direct sum of the two contributions, or it is enhanced strongly. The latter is true for CBC, AUG, TJ-II and LHD, indicating that the instabilities are destabilized significantly by the free energy in the ion temperature gradient, and by the electron dynamics. In contrast, in W7-X and NCSX, the ion heat flux from the complete simulations (black lines) resembles the direct sum of the separate effects, where NCSX benefits particularly well from the weak density-gradient-driven modes (red dotted lines).
Therefore, for W7-X and NCSX there is a region of density gradients for which the ion-temperature-gradient-driven modes are being suppressed and the density-gradient-driven modes are not driving a significant amount of turbulence yet, which is qualitatively in line with the linear observations \mbox{(figure \ref{fig:linear_teprim})}. 
Moreover, from the nonlinear simulations we can conclude that the reduction of the ion heat flux by the density gradient is also related to the fact that the ion-temperature-gradient-driven modes (blue dashed lines) in NCSX are only moderately destabilized by the electron dynamics, while they are not destabilized by kinetic electrons in W7-X, in contrast to the other devices. Additionally, the density-gradient-driven modes (red dotted lines) in NCSX and \mbox{W7-X} are slightly stabilized in the presence of a finite ion temperature gradient, while they are significantly destabilized in the other devices.

\subsection{Effect of the electron temperature gradient}
\label{sec:nonlinearWithALTe}

The effect of adding a non-vanishing electron temperature gradient on the heat transport is shown in figures \ref{fig:qi_contributions} and \ref{fig:qe_contributions}, respectively, by comparing the black lines $(a/L_{T_e}=0.0)$ with the gray lines $(a/L_{T_e}=3.0)$. It can be seen that a finite electron temperature gradient has a very small effect on the ion heat losses \mbox{(figure \ref{fig:qi_contributions})}. Specifically, in AUG and TJ-II, the presence of an electron temperature gradient  decreases the ion heat flux moderately for $a/L_n\leq 2.0$ and increases it for $a/L_n\geq 2.0$, a similar increase in the ion heat flux is observed for CBC. In W7-X, the electron temperature gradient has a small stabilizing effect on the ion heat loses.
On the other hand, the electron temperature gradient increases the drive for the electron heat transport strongly, resulting in a significant increase of the electron heat flux, which becomes of the order of the ion heat flux. Note that the electron heat flux in the absence of an electron temperature gradient (black lines in figure \ref{fig:qe_contributions}) is much smaller than the ion heat flux (black lines in figure \ref{fig:qi_contributions}), especially at low density gradients. The trend of the electron heat flux is similar when considering $a/L_{T_e}=0.0$ or $a/L_{T_e}=3.0$.
Specifically, in CBC and AUG, both the electron heat flux and the ion heat flux increase with the density gradient. For LHD, the electron heat flux increases roughly linearly with the density gradient, similar as its linear growth rates. In TJ-II the electron heat flux initially decreases, after which it increases strongly with the density gradient for $a/L_{T_e}=3.0$, comparable to the evolution of its ion heat flux. Finally, for both NCSX and W7-X, the electron heat flux undergoes a similar reduction with increasing density gradients as the ion heat flux, albeit it is less pronounced.

Recall that these simulations are performed at ion-Larmor scales, therefore the contribution to the heat transport that arises from the electron-Larmor scales is missing. Nonetheless, it has been shown that when the ion temperature gradient is sufficiently above the critical ion temperature gradient, the ion and electron heat flux driven at ion-Larmor scales gives an accurate picture of the heat transport in tokamaks \cite{howard2016multi, howard2015multi}. Whether or not this holds for stellarators needs to be investigated and is beyond the scope of this work.

\subsection{Contribution of trapped and passing particles}
\label{sec:distributon}

\begin{figure}[h!]
	\centering
	\includegraphics[trim={0mm 2mm 0mm 3mm}] {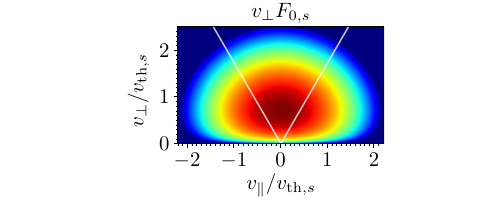}   
	\caption{Maxwellian distribution in velocity space, with a trapping cone corresponding to $B/B_\textrm{max}=0.75$.}	
	\label{fig:hmuvpa_max} 
\end{figure}

In order to gain a better understanding of the micro-instabilities that drive the turbulence, the turbulent part of the distribution function is examined in detail in this section.
Let us denote by $\hat{g}_{\mathbf{k},s}$ the Fourier components of the gyro-averaged turbulent distribution function and let  
\begin{equation}
	\hat{h}_{\mathbf{k},s} = \hat{g}_{\mathbf{k},s} +  \frac{Z_s e}{T_s} \, \hat{\varphi}_\mathbf{k}\, F_{0,s} \, J_0(k_\perp\rho_s)
	\label{eq:g}
\end{equation}
be its non-adiabatic component, where $F_{0,s}$ is a Maxwellian distribution. In figure \ref{fig:hmuvpa_max} a Maxwellian distribution is shown as a function of the parallel and perpendicular velocities. The factor $v_\perp$ in this plot corresponds to the Jacobian of the transformation from $(v_x,v_y,v_z)$ to $(v_\perp, v_\parallel)$ velocity coordinates.  
Particles are trapped if $\smash{v_\parallel^2 + v_{\!\perp}^2 < 2\mu B_\text{max}}$, with \mbox{$v_{\!\perp}^2 = 2\mu B(z)$} and $B_\text{max}$ the maximum value of $B(z)$ along the field line. This trapping condition is visualized in figure \ref{fig:hmuvpa_max} by the white lines---commonly known as the trapping cone. Passing particles (outside the cone) are capable of exploring the entire field line, while trapped particles (inside the cone) are confined to the magnetic wells. For a Maxwellian distribution, the fluctuations are isotropic in velocity space, and the trapped particle fraction is given by \mbox{$f_\text{trapped}(z) = \sqrt{1-B(z)/B_\textrm{max}}$}. In figure \ref{fig:hmuvpa_max}, we have $f_\text{trapped}=0.5$, therefore, half of the particles are trapped in this scenario.

\hfill

Note that a large fraction of trapped particles does not necessarily imply that trapping plays an important role on the turbulence,  it may simply indicate that the magnetic well is relatively deep. Only if plots of \smash{$\hat{h}_{\mathbf{k},s} (v_\parallel,v_\perp)$} exhibit a clear deviation from isotropy, can we expect that trapping could play an important role. Moreover, \smash{$\hat{h}_{\mathbf{k},s}$} has to be sufficiently large compared to \smash{$\hat{g}_{\mathbf{k},s}$} to have a notable influence on the turbulence. Therefore, it is not sufficient to only consider the fraction of trapped particles to discuss the nature of the turbulence, instead, one has to carefully examine the distribution in velocity space while simultaneously taking into account the relative size of $\hat{h}_{\mathbf{k},s}$, which is discussed in the following paragraphs.

\begin{figure}[b!]
	\centering 
	\includegraphics[trim={0mm 4mm 0mm 0mm}] {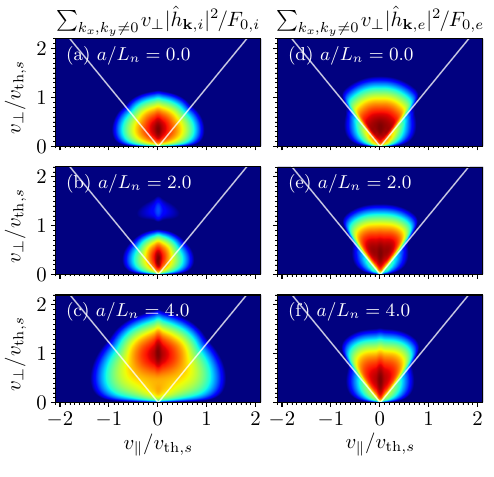}  
	\caption{Non-adiabatic part of the distribution function at \mbox{$z=0$} for ions (left) and electrons (right) in CBC considering \mbox{$a/L_{T_i}=3.0$} and $a/L_{T_e}=0.0$, for $a/L_n=0.0$ (top row),  $a/L_n=2.0$ (middle row) and  $a/L_n=4.0$ (bottom row). \vspace{2mm}}	
	\label{fig:hmuvpa_CBC}
\end{figure}

\begin{figure}[b!]
	\centering
	\includegraphics[trim={0mm 4mm 0mm 1mm}] {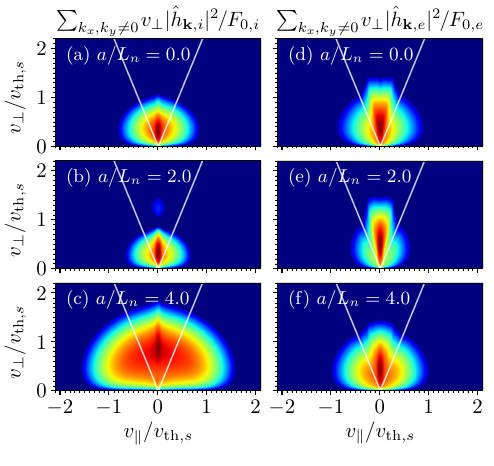}   
	\caption{Non-adiabatic part of the distribution function at \mbox{$z=0$} for ions (left) and electrons (right) in W7-X considering \mbox{$a/L_{T_i}=3.0$} and $a/L_{T_e}=0.0$, for $a/L_n=0.0$ (top row),  $a/L_n=2.0$ (middle row) and  $a/L_n=4.0$ (bottom row).}	
	\label{fig:hmuvpa_W7X} 
\end{figure}

\newpage
In figures \ref{fig:hmuvpa_CBC} and \ref{fig:hmuvpa_W7X}, the non-zonal contributions to the non-adiabatic part of the distribution function for the ions and electrons, i.e.,  $\smash{\sum_{k_x, k_y\neq 0}|\hat{h}_{\mathbf{k},s}|^2}$, are shown for CBC and \mbox{W7-X} at $z=0$, considering \mbox{$a/L_{T_i}=3.0$} and $a/L_{T_e}=0.0$. It is clear that for electrons trapping can play a dominant role, e.g., in CBC the distribution function is clearly localized within the trapping region (right column of figure  \ref{fig:hmuvpa_CBC}) and the electrons are strongly trapped for all density gradients. On the other hand, in W7-X the electrons are fairly trapped at $a/L_n=0.0$, they become very trapped at $a/L_n=2.0$ and finally they are moderately trapped at $a/L_n=4.0$ (right column of figure  \ref{fig:hmuvpa_W7X}). Note that in stellarators particles can also be trapped with respect to a local maximum of the magnetic field, which results in a more narrow trapping cone. This is observed for W7-X (right column of figure  \ref{fig:hmuvpa_W7X}), which has a shallow magnetic well at the center of the field line (figure \ref{fig:fieldline}a). On the other hand, for CBC and W7-X, ions are more strongly trapped than in the case of an isotropic distribution (figure \ref{fig:hmuvpa_max}), which can be seen in the left columns of figures \ref{fig:hmuvpa_CBC} and \ref{fig:hmuvpa_W7X}. Except for $a/L_n=4.0$ in W7-X, where the ion distribution is almost isotropic. Note however that trapping does not necessarily play a notable role on the ion dynamics, it is possible that the instability is simply predominantly driven by ions that have small parallel velocities. This could explain why the ion distribution is rather smooth across the trapping-passing boundary, while for electrons the distribution changes strongly across the trapping-passing boundary.  

\begin{figure}[b!]
	\centering   
	\includegraphics[trim={0mm 1mm 0mm 2mm}] {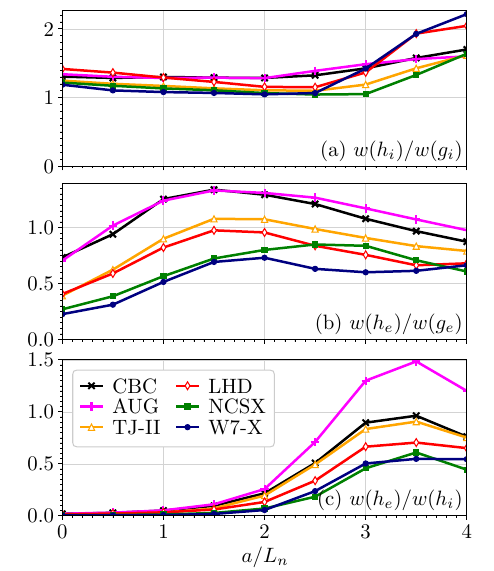} 
	\caption{Ratio of the non-adiabatic part of the distribution function to the gyro-averaged distributed function (a) for ions and (b) for electrons, and (c) the fraction of the non-adiabatic part of the distribution function of the electrons to the ions, considering $a/L_{T_i}=3.0$ and $a/L_{T_e}=0.0$.}
	\label{fig:ratios}
\end{figure}

To quantify the degree of non-adiabaticity and the importance of trapped particles we are going to introduce new diagnostics in the following paragraphs. Let us define
\begin{equation}
	\label{eq:deltan}
	w(h_s, z) = \int d^3\mathbf{v} \!\! \sum\limits_{k_x, k_y \neq 0}  \frac{T_s}{2} \frac{|\hat{h}_{\mathbf{k},s}|^2}{F_{0,s}} .
\end{equation}
Given that the role of trapped particles could be significant in $h_s$, but, at the same time, the non-adiabatic part $h_s$ could be very small compared to the distribution $g_s$ in some situations, we quantify the degree of non-adiabaticity as $w(h_i)/w(g_i)$ and $w(h_e)/w(g_e)$, which are shown in figures \ref{fig:ratios}a and \ref{fig:ratios}b, respectively. Note that $w$ (with dimensions of energy density) is reminiscent to one of the pieces of the free energy (see equation (74) in \cite{schekochihin2009astrophysical}). This quantity is evaluated at $z=0$ for CBC, AUG, NCSX and \mbox{W7-X}, and at $z=\pm 0.23$ for LHD and $z = \pm 0.54$ for \mbox{TJ-II} since these are the locations where the potential fluctuations peak. The parallel mode structures of the potential are not shown here, nonetheless, they are similar to the localization of the heat losses shown in figure \ref{fig:qfluxvsz}, with the exception that the potential fluctuations always peak at the center for W7-X, and in the deepest wells for LHD and \mbox{TJ-II}, regardless of the density gradient. The parameter $z$ is henceforward dropped in $w(h_s, z)$ for simplicity.

For the ions $w(h_i)>w(g_i)$ for all density gradients in every device (figure \ref{fig:ratios}a), which means that the potential counter-acts the ion dynamics. Moreover, the ratio is around unity for small density gradients, and it increases substantially at large density gradients, hence the ion dynamics are almost entirely non-adiabatic. On the other hand, for small density gradients, the contribution of the electrons is much smaller than that of the ions, i.e., \mbox{$w(h_e) \ll w(h_i)$} (figure \ref{fig:ratios}c), moreover, the non-adiabatic part of the electrons is rather small, $w(h_e) < w(g_e)$ \mbox{(figure \ref{fig:ratios}b)}.  Therefore, for small density gradients, the electron dynamics do not play a significant role on the turbulence. 
As the density gradient is increased from \mbox{$a/L_n=0.0$} to $a/L_n=1.5$, the ITG mode is damped, and while the electron dynamics remain negligible ($w(h_e) \ll w(h_i)$), the non-adiabatic part of the electron distribution function (figure \ref{fig:ratios}b) increases with the density gradient. When the density gradient is increased further, $a/L_n\geq2.0$, the electron dynamics start playing an important role since $w(h_e)/w(h_i)$ becomes relatively large, reaching values around 0.5 for W7-X and NCSX, up to 1.5 in AUG (figure \ref{fig:ratios}c). Moreover, the non-adiabatic part of the distribution function of the electrons is relatively large at $a/L_n=2.0$, with  $w(h_e) > w(g_e)$ in  CBC and AUG, $w(h_e) \approx w(g_e)$ in TJ-II and LHD, while $w(h_e) < w(g_e)$ in W7-X and NCSX. After $a/L_n=2.0$, $w(h_e)/w(g_e)$ decreases slowly with the density gradient in CBC, AUG, TJ-II and LHD, whereas in NCSX this decrease starts after $a/L_n=3.0$ and in W7-X the decrease starts at $a/L_n=2.0$, however $w(h_e)/w(g_e)$ starts increasing again after $a/L_n=3.0$. 

\begin{figure}[b!]
	\centering   
	\includegraphics[trim={0mm 1mm 0mm -2mm}] {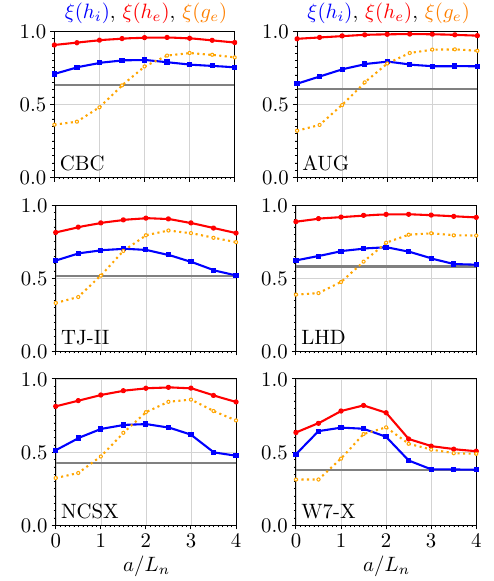}  
	\caption{ Fraction of trapped particles considering $h_i$ (blue), $h_e$ (red) and $g_e$ (orange), for $a/L_{T_i}=3$,  $a/L_{T_e}=0$, evaluated at the center of the field line $(z=0)$ for CBC, AUG, NCSX and W7-X and at $z=\pm 0.23$ for LHD and  $z = \pm 0.54$ for \mbox{TJ-II}. The trapped particle fraction $f_\text{trapped} = \sqrt{1-B(z)/B_\textrm{max}}$ is visualized by the horizontal gray lines. \vspace{2mm}}
	\label{fig:trappedratios}
\end{figure}

To quantify whether the turbulence is driven by trapped or passing particles, we define a measure of the particle trapping at each point $z$ as, 
\begin{equation}
	\label{eq:deltan}
	\xi(h_s, z) = \frac{1}{w(h_s, z)} \int  \! d^3\mathbf{v} \ H_\text{trap} \!\!\! \sum\limits_{k_x, k_y \neq 0}  \frac{T_s}{2} \frac{|\hat{h}_{\mathbf{k},s}|^2}{F_{0,s}} ,
\end{equation}
with $H_\text{trap} = H(2\mu (B_\text{max}\!-\!B(z)) - v_\parallel^2)$ the Heaviside function. This measure is evaluated at the same $z$-locations considered in the previous paragraph, hence $z$ is dropped in $\xi (h_s,z)$ for simplicity. It is plotted in figure \ref{fig:trappedratios} considering the non-adiabatic part of the distribution function for the ions ($h_i$ in red) and the electrons ($h_e$ in blue), as well as for the full turbulent distribution function of the electrons ($g_e$ in orange). 

The quantity $\xi$ needs to be compared to the fraction of trapped particles that would occur if the fluctuations were isotropic in velocity space, which corresponds to the trapped particle fraction \mbox{$f_\text{trapped} = \sqrt{1-B(z)/B_\textrm{max}}$}, represented by the gray lines in figure \ref{fig:trappedratios}. The trapped particle fraction is 63\% in CBC, 60\% in AUG, 51\% in TJ-II, 58\% in LHD, 43\% in NCSX and 38\% in W7-X.
For small density gradients, the majority of the ions are trapped, i.e., $\xi(h_i) > f_\text{trapped}$ (blue lines in figure \ref{fig:trappedratios}).   As the density gradient is increased, $\xi(h_i)$ approaches $f_\text{trapped}$  in the stellarators (blue lines $\rightarrow$ gray lines), because the velocity distribution becomes isotropic. In contracst, in CBC and AUG $\xi(h_i)$ remains high at $a/L_n=4.0$. 

On the other hand, the majority of the electrons which are driving the non-adiabatic part of the distribution function (red lines) are trapped regardless of the density gradient, i.e., $\xi(h_e) > f_\text{trapped}$. With the exception of W7-X, which will be discussed separately in the next section. The non-adiabatic part $h_e$ contains the electrons which are driven unstable by the geometric characteristics, whereas the distribution function $g_e$ also contains the contribution of the electrons that follow the perturbed electrostatic potential $\varphi$. Therefore, the geometric characteristics, such as the magnetic wells and regions of bad curvature, mostly excite trapped electrons, hence $\xi(h_e)$ is close to unity (red lines). 
Finally, looking at the full distribution function of the electrons (orange lines), it is clear that for flat density profiles the electrons are mostly passing, $\xi(g_e) < f_\text{trapped}$. This is because the electron dynamics do not play a notable role at small density gradients (figure \ref{fig:ratios}) and hence the majority of the electrons are passing since they respond to the electrostatic potential determined by the ion dynamics. As the density gradient is increased, and the electron dynamics start playing a significant role, the majority of the electrons become trapped, $\xi(g_e) > f_\text{trapped}$, in CBC, AUG, TJ-II, LHD and NCSX where the trapped-electron-mode is believed to become dominant.

\subsection{The case of W7-X: trapped-electron-modes and passing-particle-driven universal instabilities}

The amount of trapped and passing particles that are driving the turbulence in W7-X is very different. Specifically, $\xi(h_e)$ is significantly lower in \mbox{W7-X} compared to the other five devices (red lines in \mbox{figure \ref{fig:trappedratios}}), and it depends strongly on the density gradient.  In particular, for small density gradients, the particles that are driving the instability become increasingly trapped as the role of the electron dynamics increases (figures \ref{fig:ratios}b \mbox{and \ref{fig:ratios}c}), similarly as in the other devices. 
On the other hand, for \mbox{$a/L_n \geq 2.5$}, $\xi(h_e)$  decreases significantly, indicating that the turbulence is driven by both trapped and passing electrons. Therefore, in \mbox{W7-X, the} density-gradient-driven modes likely consist of the trapped-electron-modes, as well as another instability that is driven by passing electrons, which is most likely the passing-particle-driven \mbox{universal instability \cite{costello2023universal}}. This is supported by figure \ref{fig:hmuvpa_W7X}f which shows that the non-adiabatic part of the distribution function consists of a trapped particle contribution coming from the central magnetic well (since the cone is very narrow) with the addition of an isotropic contribution which is likely originating from the passing-particle-driven universal instability. Moreover, this instability explains the large amount of heat loss across the entire field line for \mbox{$a/L_n \geq 2.5$} in W7-X (figure \ref{fig:qfluxvsz}), since passing particles explore the entire geometry. Comparing with the results in \cite{costello2023universal} this instability likely corresponds to instability (F) in the linear simulations shown in figure \ref{fig:gamma_map_line}.  
Therefore, it is likely that in W7-X the trapped-electron-mode co-exists with the passing-particle-driven universal instability at high density gradients. In the other five devices, the passing-particle-driven universal instability is probably absent, or sub-dominant to the trapped-electron-mode. The detailed analysis of the distribution function thus gives valuable insights into the micro-instabilities which are driving the turbulence.

\section{Conclusions}
\label{sec:conclusions}

The effect of the density gradient on the turbulent heat transport is investigated by carrying out a comprehensive comparative study of different devices, including the  W7-X, LHD, TJ-II and NCSX stellarators, which represent helias, heliotron, heliac and quasi-axisymmetric configurations, respectively, as well as the Asdex-Upgrade (AUG) tokamak and the tokamak geometry of the Cyclone Base Case (CBC). Both linear and nonlinear simulations covering a wide range of parameters are performed, examining the effect of density gradients, as well as the presence of electrons with and without a temperature gradient. Moreover, a detailed study of the geometric characteristics---magnetic wells, curvature and shear---that drive the instabilities is conducted, and the contributions of trapped and passing particles to the turbulence  are identified. 
 
A detailed study of the linear growth rates along $(k_y\rho_i, a/L_n)$ reveals that many distinct instability branches co-exist at ion-scales in \mbox{W7-X}, while only two branches co-exist in NCSX, which can have an important effect on the nonlinear dynamics, hence one has to be careful when studying only the most unstable mode along $k_y\rho_i$. On the other hand, in the other devices a continuous transition between different micro-instabilities has been observed for most of the considered density gradients, ion temperature gradients, and wavenumbers, hence the identification of the instabilities should be done very carefully or be avoided altogether. It has been shown that the growth rates increase with increasing density gradients in CBC, AUG, TJ-II, LHD and NCSX, while a moderate decrease of the growth rate is found between $a/L_n = 1.25$ and $a/L_n = 2.75$ in \mbox{W7-X}. Moreover, in NCSX and W7-X the growth rates of the density-gradient-driven modes exceeds those driven by the ion-temperature-gradient-driven modes around \mbox{$a/L_n \approx 2.0$}, whereas this occurs around $a/L_n \approx 1.0$ for CBC, AUG, TJ-II and LHD. Therefore, in NCSX and W7-X there is a region of density gradients for which the ion-temperature-gradient-driven modes are stabilized, without driving the density-gradient-driven modes particularly unstable yet.



\begin{table*}[b!]
	\centering
	\vspace{2mm}
	\caption{\label{tab:resolution} Resolution used for the linear and nonlinear simulations included in this paper. For nonlinear simulations, approximately 1 poloidal turn is used to ensure that $\Delta k_x = \Delta k_y$, the exact length of the flux tube is given in table \ref{tab:geometry}. For LHD, $N_z=97$ grid points are used in nonlinear simulations to account for its more detailed magnetic field structure. The Fourier grid corresponds to a computational domain of $L_x = L_y = 94\rho_i$ with $N_x = N_y = 91$ grid points in real space, here $\rho_i$ is the ion-Larmor radius.} \vspace{-4mm} 
	\lineup
	\begin{indented}
		\item[]\begin{tabular}{@{}ccccccccccc}
			\br & \centre{2}{Parallel grid}
			& \centre{4}{Fourier grid} & \centre{4}{Velocity grid}  \\[-1.5mm]
			& \crule{2} & \crule{4} & \crule{4} \\[0mm]
			& $N_{\rm pol}$ & $N_z$ & $k_x\rho_i$ & $k_y\rho_i$ & $N_{k_x}$ & $N_{k_y}$ & $v_\parallel/v_{{\rm th},s}$ & $v_\perp/v_{{\rm th},s}$ & $N_{v_\parallel}$& $N_{v_\perp}$  \\
			\qquad&\qquad&\qquad&
			\qquad\quad\;\;\;\; &\qquad\quad\;\;\;\; &\qquad&\qquad&
			\qquad\quad\;\;\; &\qquad\quad\;\;\; &\qquad&\qquad\\[-4mm]
			\mr
			Linear & 3 & 513  & 0 & [0.125, 2.00] & 0 & $>$20 & [-3.0, 3.0] & [0.0, 3.0] & 128 & 24   \\[1mm]
			Nonlinear & $\sim$1 & 49  & [0.067, 2.00] & [0.067, 2.00] & 61 & 31  & [-3.0, 3.0] & [0.0, 3.0] & 48& 12    \\ 
			\br
		\end{tabular}
	\end{indented} 
	\vspace{3mm}
\end{table*}

The analysis is continued by performing nonlinear simulations covering a wide range of density gradients, $0.0 \leq a/L_n \leq 4.0$, considering a fixed \mbox{ion temperature} gradient of $a/L_{T_i}=3.0$ and taken into account both vanishing and finite electron temperature gradients. In NCSX and W7-X, a strong reduction of the ion heat flux with increasing density gradients is observed. This reduction is also evident in TJ-II for $a/L_n \leq 1.0$, whereas in LHD, the ion heat flux is nearly unaffected by the density gradient.  In contrast, in the tokamaks, the ion heat flux increases strongly with the density gradient. Notably, the ion heat fluxes for the stellarators do not correlate with the growth rates obtained in the linear analysis. 
The strong reduction of the ion heat flux in \mbox{W7-X} and NCSX is attributed to the stabilization of the ion-temperature-gradient-driven modes by the density gradient, without significantly destabilizing the \mbox{density-gradient-driven modes.} Moreover, NCSX and W7-X benefit from the fact that the ion-temperature-gradient-driven modes are only moderately or not at all destabilized by kinetic electrons, while the density-gradient-driven modes are slightly stabilized in the presence of an ion temperature gradient. This is in contrast to the other four devices where these modes are significantly destabilized by kinetic electrons and ion temperature gradients. Furthermore, in NCSX, W7-X and TJ-II the reduction of the ion heat flux can be attributed in part to the fact that the size of the perpendicular fluctuations diminishes as the density gradient increases.
Lastly, more light is shed on the nature of the micro-instabilities by investigating the contributions of the trapped and passing particles to the turbulence. This reveals that trapped-electron-modes are dominant at high density gradients in CBC, AUG, TJ-II, LHD and NCSX, while the passing-particle-driven universal instability drives a large part of the turbulence in W7-X.

\begin{figure*}[b!]
	\centering 
	\includegraphics[trim={0mm 1mm 0mm 0mm}]{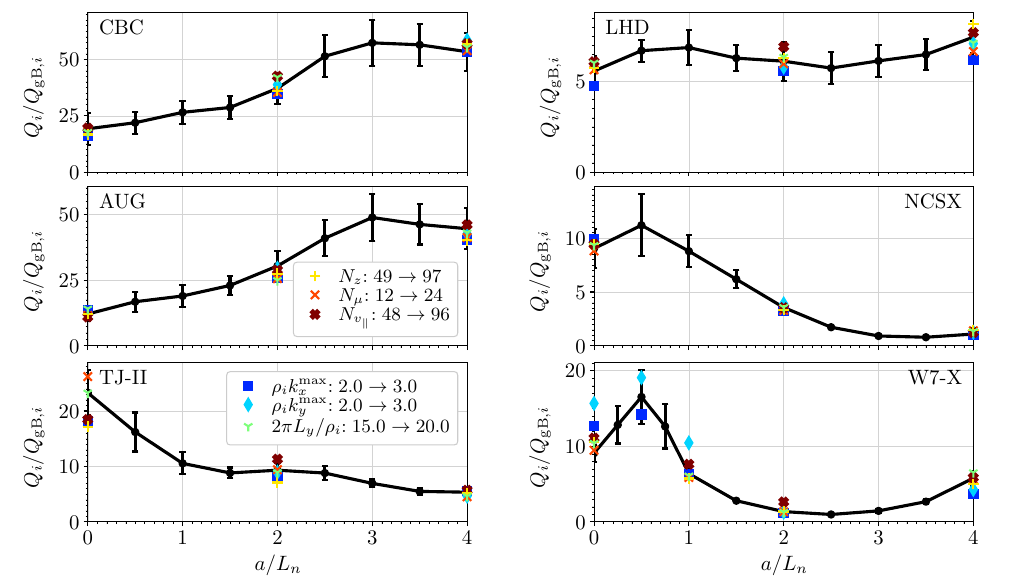} 
	\caption{Convergence checks by increasing the nonlinear resolution parameters one by one, considering $a/L_{T_i}=3.0$ and \mbox{$a/L_{T_e}=0.0$}, corresponding to the simulations in figure \ref{fig:Qi_vs_aLn}. For LHD, the parallel resolution is increased from $N_z = 97$ to $193$. }
	\label{fig:resolution}
\end{figure*} 

\ack
This work has been carried out within the framework of the EUROfusion Consortium, funded by the European Union via the Euratom Research and Training Programme (Grant Agreement No 101052200 -- EUROfusion). Views and
opinions expressed are however those of the author(s) only and do not necessarily reflect those of the European Union or the European Commission. Neither the European Union nor the European Commission can be held responsible for them. This research was supported in part by grant PID2021-123175NB-I00, Ministerio de Ciencia e Innovaci\'on, Spain. Simulations are performed on the supercomputer Marconi (CINECA, Italy).

\appendix


\section{Convergence checks}
\label{sec:resolution} 

The linear and nonlinear resolutions used in this work, are given in \mbox{table \ref{tab:resolution}}. The nonlinear resolutions have been obtained from very rigorous convergence checks performed in W7-X for different sets of density and ion temperature gradients. The same resolution has been used for the other five devices, checking the validity of the choices by doubling the resolution parameters one by one, and examining the effect on the ion heat flux, which is shown in figure \ref{fig:resolution}. Here the black lines represent the ion heat fluxes presented in figure \ref{fig:Qi_vs_aLn}. First tests for LHD revealed that $N_z=49$ is too low in order to resolve its detailed magnetic field structure (figure \ref{fig:fieldline}). Therefore, a resolution of $N_z=97$ is used for LHD and it is doubled to $N_z = 193$ in figure \ref{fig:resolution}.
 
The error bars in figure \ref{fig:resolution} represent the standard mean deviation on the time trace of the simulation, taken between $tv_\text{th,i}/a=500$ and \mbox{$tv_\text{th,i}/a=1000$}, where the heat fluxes are saturated. In the simulations run with higher resolutions (colored markers), an ion heat flux has been obtained which falls within the error bars of the original simulations, hence the resolution used in this work produces reliable heat fluxes. Only in W7-X does the resolution in the perpendicular $(k_x,k_y)$-grid change the ion heat flux beyond the error bars, nonetheless, the change is not significant and it does not influence the conclusions drawn in this paper. 

\begin{figure}[b!] 
	\centering 
	\includegraphics[width=0.9\linewidth, trim={0mm 2mm 0mm 3mm}]{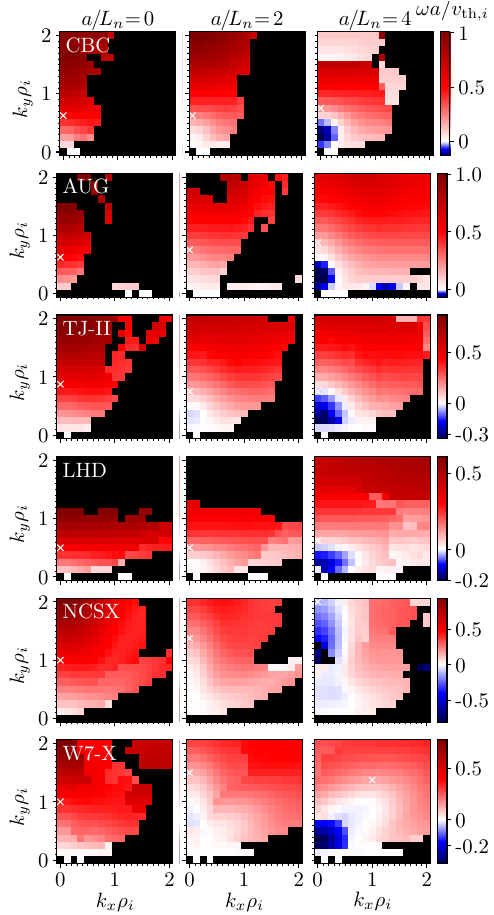} 
	\caption{Normalized frequency as a function of $(k_x\rho_i, k_y\rho_i)$ for $a/L_{T_i}=3.0$, $a/L_{T_e}=0.0$ and $a/L_n=0.0$ (left column), $a/L_n = 2.0$ (middle column) and $a/L_n=4.0$ (right column). The most unstable mode is highlighted with a white cross.}
	\label{fig:omega_kxky}
\end{figure} 

\section{Frequency as a function of the radial and binormal wavenumbers}
\label{app:omega}

In figure \ref{fig:gamma_kxky} the growth rates are shown as a function of the radial and binormal wavenumbers (section \ref{sec:gammavskxky}). The corresponding frequencies of these instabilities are shown in figure \ref{fig:omega_kxky}. For flat density profiles (left column) only positive frequencies are observed for all considered $(k_x,k_y)$-modes. When the density gradient is increased to $a/L_n=2.0$ (middle column), slightly negative frequencies appear in TJ-II and W7-X around $k_x\rho_i = 0.0$. Finally, negative frequencies are observed in every device when considering a strong density gradient of $a/L_n=4.0$ (right column).

\begin{figure*}[!b]
	\centering 
	\includegraphics[trim={0mm 0mm 0mm 0mm}] {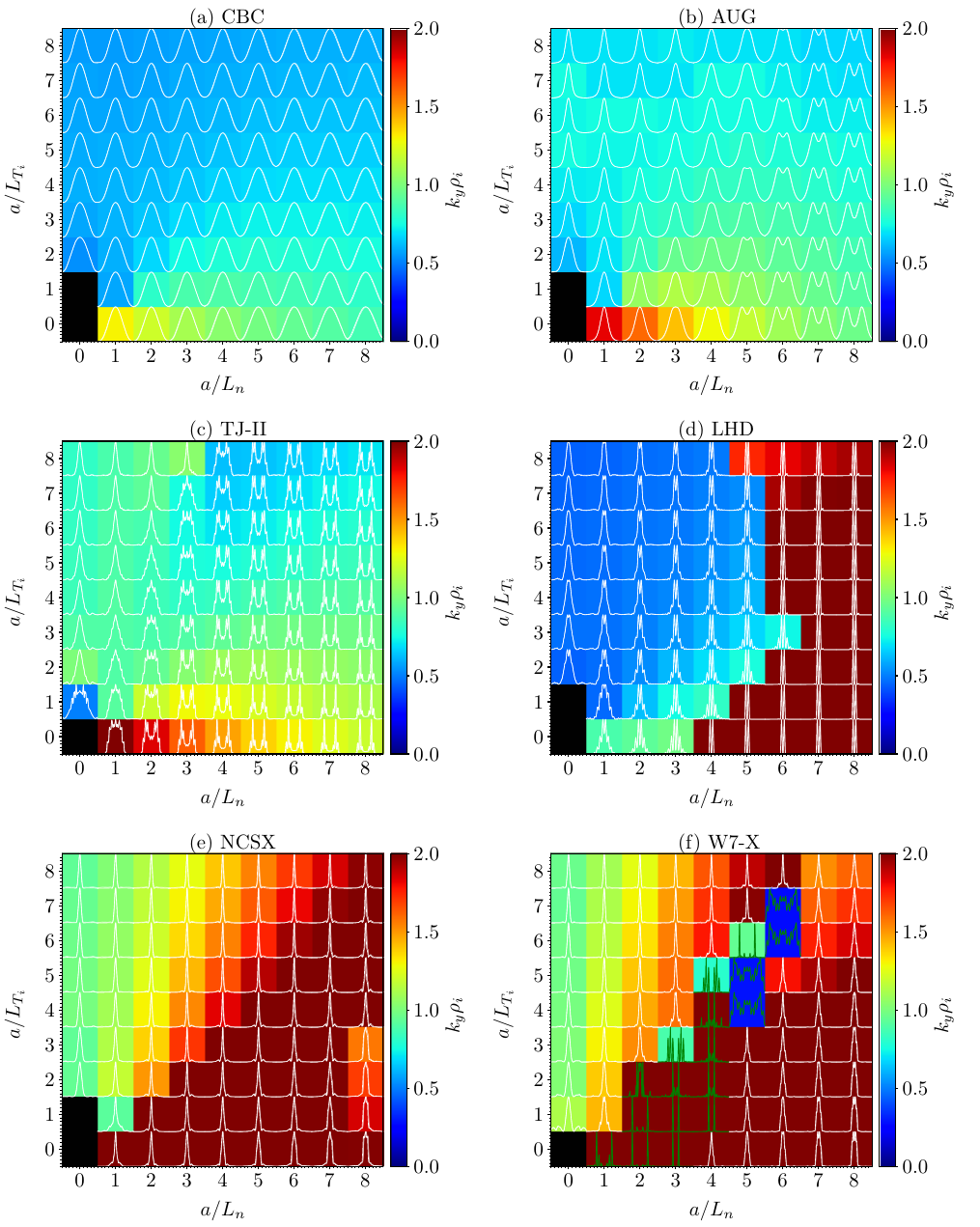}  
	\caption{Binormal wavenumber, $k_y\rho_i$,  of the most unstable mode for $k_x\rho_i=0.0$ and $k_y\rho_i \leq 2.0$, as a function of the normalized density gradient and the normalized ion temperature gradient, considering a vanishing electron temperature gradient. }
	\label{fig:ky_maps}
\end{figure*}

\section{Wavenumber of the most unstable mode as a function of the density gradient and the ion temperature gradient}
\label{app:ky} 

In section \ref{sec:gammavsfprimtiprim} the growth rates and frequencies of the most unstable mode, considering $k_x=0.0$, $k_y\leq 2.0$ and $a/L_{T_e}=0.0$, are shown in figures \ref{fig:gamma_maps} and \ref{fig:omega_maps}, respectively. The corresponding wavenumber of the most unstable mode is shown in figure \ref{fig:ky_maps}. In NCSX and  W7-X, the instabilities drive smaller perpendicular fluctuations (larger $k_y\rho_i$) as the density gradient is increased. In contrast, in CBC, AUG, TJ-II and LHD, the most unstable mode retains large perpendicular scales when increasing the density gradient. Note that for W7-X, NCSX and LHD, the most unstable modes in the bottom right corners, are found at the end of our scanned range of $k_y\rho_i$ values. Therefore, the growth rates and frequencies shown in figures  \ref{fig:gamma_maps} and \ref{fig:omega_maps} do not correspond to a peak in the $\gamma(k_y\rho_i)$ spectra. This is the reason why the frequency spectrum of NCSX has such negative frequencies in the bottom right corner. If one would increase the scanned $k_y\rho_i$ values, the most unstable mode would be found further along the frequency branch, for higher, and possibly positive, frequencies. Nonetheless, we restricted the linear studies to  $k_y\rho_i\leq 2.0$ to remain on ion scales. 

On top of each $k_y\rho_i(a/L_n, a/L_{T_i})$ tile, the parallel mode structure of the most unstable mode is shown, by plotting the potential squared along the field line, plotted in white if it is confined to the first poloidal turn, or in green if the mode extends up to three poloidal turns. It can se seen that in CBC and AUG, the binormal wavenumber of the most unstable mode and its parallel mode structure change continuously throughout the scanned gradients. Similarly, in \mbox{TJ-II} and NCSX the binormal wavenumber and parallel mode structure change continuously for most of the scanned gradients, however, for some adjoining tiles, the mode changes rather strongly. Nonetheless by going from any tile A to tile B by first passing through other adjoining tiles, the changes are continuous. When looking at the $\gamma(k_y\rho_i)$ spectra (not shown here) this corresponds to a single peak in $\gamma(k_y\rho_i)$ which slowly splits up into two peaks as the gradients are modified. Finally, for LHD and W7-X clear jumps between distinct modes can be observed. Specifically, in W7-X, extended modes with both large and small perpendicular scales show up along $a/L_n \approx a/L_{T_i}$.

\newpage
\section*{References}
\bibliographystyle{unsrt}
\bibliography{bibliography}

\end{document}